\documentclass[11pt]{article}

\usepackage{jheppub}
\usepackage[usenames,dvipsnames,table]{xcolor}
\usepackage{graphicx,amsmath,amssymb,amsthm,multirow,array,bm,bbm,esint,slashed,hyperref}
\usepackage{xfrac,faktor}
\usepackage[mathscr]{eucal}
\usepackage[bbgreekl]{mathbbol}
\usepackage{subfig}

\newcommand\beq{\begin{equation}}
\newcommand\eeq{\end{equation}}
\newcommand{\nn}{\nonumber}

\title{Chern-Simons dualities with multiple flavors at large $N$}

\author{Kristan Jensen$^a$}
\author{and Priti Patil$^{a,b}$}
\affiliation{$^a$Department of Physics and Astronomy, San Francisco State University, San Francisco, CA 94132}
\affiliation{$^b$Department of Physics, University of California, Davis, CA 95616}

\preprint{\today}

\emailAdd{kristanj@sfsu.edu}
\emailAdd{ppatil@ucdavis.edu}

\abstract{We study $U(N)_k$ Chern-Simons theory coupled to fundamental fermions and scalars in a large $N$ `t Hooft limit. We compute the thermal free energy at high temperature, as well as two- and three-point functions of simple gauge-invariant operators. Our findings support various dualities between Chern-Simons-matter theories with $\mathcal{N}=0,1,$ and $2$ supersymmetry.
}

\begin{document}
\maketitle
%***********************************************

%***********************************************
\section{Introduction}
%***********************************************

Large $N$ quantum field theories with matrix fields are generally insoluble. One important exception is that of Chern-Simons matter theories in an `t Hooft limit. These theories give us a theoretical lamppost, under which we can reliably compute many observables, like the $S$-matrix of massive phases~\cite{Jain:2014nza,Inbasekar:2015tsa,Inbasekar:2017ieo}, operator product expansion (OPE) coefficients at fixed points~\cite{Aharony:2011jz,Giombi:2011kc,Aharony:2012nh,GurAri:2012is}, finite temperature response functions~\cite{Aharony:2012ns,Jain:2013py,Geracie:2015drf,Gur-Ari:2016xff}, and so on. 

One of the most interesting insights from these large $N$ computations (of which there are many more e.g.~\cite{Shenker:2011zf,Jain:2012qi,Jain:2013gza,Gurucharan:2014cva,Bedhotiya:2015uga,Gur-Ari:2015pca,Giombi:2016zwa,Giombi:2017rhm,Charan:2017jyc,Turiaci:2018dht,Choudhury:2018iwf,Aharony:2018pjn,Skvortsov:2018uru,Dey:2019ihe,Aharony:2019mbc,Jain:2019fja}) is that Chern-Simons matter theories exhibit dualities even in the absence of supersymmetry (SUSY). It had been well-known that pure Chern-Simons theories enjoy level/rank dualities, for example equating Chern-Simons theory with gauge group $U(N)$ and level $k$ to one with gauge group $SU(k)$ and level $-N$~\cite{Naculich:1990pa,Mlawer:1990uv,Nakanishi:1990hj} (see also~\cite{Hsin:2016blu}). In a sense this duality survives when both gauge theories are appropriately coupled to a single fundamental field at large $N$, $k$ with $N/k$ held fixed. The basic example relates Chern-Simons theory coupled to a Dirac fermion, to Chern-Simons theory coupled to a scalar with a mass and quartic coupling tuned to a Wilson-Fisher like critical point.

In the years since there has been a significant amount of evidence accumulated that there is a wide family of these dualities, and that they persist to finite $N$, even down to $N=1$. In this article we focus on Chern-Simons theory coupled to matter in the fundamental representation. In the absence of SUSY, there are several broad sequences of proposed dualities. The first is the ``master duality'' of~\cite{Benini:2017aed,Jensen:2017bjo}, which relates Chern-Simons gauge theory coupled to fundamental Dirac fermions $\psi$ (we label the fermion of the dual as $\Psi$) and complex ``Wilson-Fisher'' scalars $\phi$ ($\Phi$); its precise statement at finite $N$ and $k$ is
\beq
\label{E:master}
	SU(k)_{-N+ \frac{N_f}{2}} \text{ with } N_f \, \psi\,, N_s \, \phi \quad \leftrightarrow \quad U(N)_{k-\frac{N_s}{2}}  \text{ with }N_s \, \Psi\,, N_f \, \Phi\,,
\eeq
provided that the ``flavor bounds'' $N_f \leq N$, $N_s \leq k$, and $(N_f, N_s) \neq (N,k)$ hold. Here the subscript on the gauge group indicates the Chern-Simons level. There are also versions with unitary gauge groups on both sides that follow from this proposal, as well as proposals with orthogonal and symplectic gauge groups. The previously proposed dualities of~\cite{Aharony:2015mjs,Aharony:2016jvv} are a special case of~\eqref{E:master}, with either $N_f = 0$ or $N_s =0$.\footnote{When either $N_f$ or $N_s = 0$, i.e. when we have Chern-Simons theory coupled to purely scalar or fermion matter, there is evidence~\cite{Komargodski:2017keh} for a different dual description slightly beyond the ``flavor bounds.''} For other works at finite $N$ see~\cite{Radicevic:2015yla,Karch:2016aux,Jensen:2017dso,Gomis:2017ixy,Aitken:2018joi,Aitken:2019shs}, including analogous proposals for dualities between quiver gauge theories. There has also been progress in embedding these dualities into string theory and AdS/CFT~\cite{Jensen:2017xbs,Armoni:2017jkl,Aitken:2018cvh}. 

There are significant checks of this proposal in the `t Hooft limit when there is a single scalar or fermion on either side. In that limit the gauge-invariant operators are the singlet operators $|\phi|^2$ or $\bar{\psi} \psi$ of dimension $\Delta = 2 + O(1/N)$, a $U(1)$ conserved current, the stress tensor, and a tower of approximately conserved currents of spin $s>2$. One important technical simplification is that, in this limit, there are no approximately marginal operators.\footnote{\label{FN:master}Both sides of the proposed duality~\eqref{E:master} are characterized by discrete data, the rank $N$, level $k$, and the number of fundamental fields, as well as an approximately marginal operator $(\bar{\psi}\phi)(\phi^{\dagger}\psi)$. It is not known whether this operator becomes irrelevant or relevant at large but finite $N$. %However we note that, when both $N_f$ and $N_s$ are nonzero, the proposal~\eqref{E:master} relies on the dynamics of this operator. In order for the semiclassical phase diagram of one side to match the other, in the Higgs phase there must be a dynamically generated mass of a particular sign for the fermions which are singlets under the unbroken gauge group~\cite{Benini:2017aed,Jensen:2017bjo}.
} 

Away from the `t Hooft limit it is impossible to perform concrete computational checks that observables, say the thermal free energy, are the same on one side of the duality as the other. However there are some powerful, non-perturbative, albeit indirect checks of the proposal. These include the matching of `t Hooft anomalies, the matching of massive semiclassical phases, and the matching of quantum numbers of baryon operators on the $SU$ side with those of monopole operators on the $U$ side. These checks build upon those already performed in~\cite{Aharony:2015mjs,Aharony:2016jvv,Benini:2017dus} when each side of the duality has only fundamental fermions or scalars. In the latter case, the simplest possible version of the duality is at $N=k=1$, equating a free Dirac fermion with $U(1)_1$ Chern-Simons theory coupled to a Wilson-Fisher scalar. Assuming that this duality holds, the authors of~\cite{Karch:2016sxi,Seiberg:2016gmd} (see also~\cite{Murugan:2016zal}) have derived an entire web of dualities using path integral techniques. This web includes particle/vortex duality, which has been long-established, giving another indirect piece of evidence.

There are also strong arguments from a lattice construction~\cite{Chen:2017lkr,Chen:2018vmz} for the dualities~\eqref{E:master} when $N_s = 0$ for $k=1$ and general $N_f,N$ satisfying $N_f \leq N$. Finally, there is some reason to believe that the dualities~\eqref{E:master}, at least for certain values of the parameters, are inherited from parent dualities between supersymmetric theories with $\mathcal{N}=2$ SUSY~\cite{Gur-Ari:2015pca,Kachru:2016rui,Kachru:2016aon}.

Dualities of the sort~\eqref{E:master} are often called ``three-dimensional bosonization'' because they exchange fermions with bosons. Such an exchange is familiar from the physics of flux attachment. Roughly speaking, many of these Chern-Simons dualities are the non-abelian version of flux attachment in relativistic quantum field theory.

A second broad class of proposed dualities concerns Chern-Simons-matter theories in the `t Hooft limit where the matter is somewhat different than studied above. At large $N$ and $k$, the proposal reads
\beq
	SU(k)_{-N } \text{ with } \,\psi \quad \approx \quad U(N)_k \text{ with } \Phi\,,
\eeq
where now $\psi$ is a ``critical'' Dirac fermion, the analogue of a Wilson-Fisher scalar, and $\Phi$ is an ordinary complex scalar. At large $N$ the lowest-dimension gauge-invariant operator on both sides is $\bar{\psi}\psi$ or $|\Phi|^2$, both of dimension $\Delta = 1 + O(1/N)$. Consequently both sides are characterized by an approximately marginal sextic coupling, and the duality map not only exchanges level and rank but also maps the sextic couplings. The beta function of the sextic coupling has recently been computed to $O(1/N)$~\cite{Aharony:2018pjn}, with the result that the approximate line of fixed points becomes, on each side, three isolated fixed points. There is a consistent duality map between the three fixed points of the $SU$ theory with the three fixed points of the $U$ theory. Less is known about these theories at finite $N$.

A third class involves Chern-Simons-matter theories where, on both sides, there are ``regular'' fundamental fermions $\psi$ and scalars $\phi$. Let $N_s$ and $N_f$ denote the number of scalars and fermions of one side, and $N_s'$ and $N_f'$ the number of the other. These dualities differ a bit from those above. In the `t Hooft limit the dimension of $|\phi|^2$ is $1 + O(1/N)$ and the dimension of $\bar{\psi}\psi$ is $2 + O(1/N)$. Consequently any such dual pair must $N_s=N_s'$ and $N_f=N_f'$, with the duality map taking scalar (fermion) bilinears to scalar (fermion) bilinears.

The best understood dualities of this type are the Giveon-Kutasov dualities~\cite{Giveon:2008zn} (and their more general cousins~\cite{Benini:2011mf}) between $\mathcal{N}=2$ SUSY Chern-Simons-matter theories. The simplest models in question here are Chern-Simons gauge theories coupled to $N_f$ chiral multiplets in the fundamental representation with a superpotential tuned to the unique point preserving $\mathcal{N}=2$ supersymmetry. These theories are completely specified by the gauge group, level, and matter content. For $U(N)_k$ Chern-Simons theory coupled to a single chiral multiplet there is evidence for the duality at large $N$ stemming from direct computation~\cite{Aharony:2012ns,Jain:2013gza}, and at finite $N$ there are checks from the computation of sphere partition functions via supersymmetric localization~\cite{Benini:2011mf} as well as brane constructions~\cite{Giveon:2008zn,Benini:2011mf}.

These isolated $\mathcal{N}=2$ points are, at large $N$, connected to a manifold of $\mathcal{N}=1$ theories parameterized by approximately marginal superpotential deformations. These $\mathcal{N}=1$ theories have been studied very recently~\cite{Aharony:2019mbc} for $U(N)_k$ gauge group coupled to $N_f$ chiral multiplets. One complication that arises now relative to the examples mentioned above is that there are two approximately marginal couplings (consistent with the supersymmetry) with different flavor structure under the $SU(N_f)$ global symmetry that rotates the $N_f$ chiral multiplets. 

Going all the way down to $\mathcal{N}=0$ supersymmetry, the only studies of these Chern-Simons dualities to date is in the `t Hooft limit with a single fermion and scalar on both sides, including the computation of the thermal free energy~\cite{Aharony:2012ns,Jain:2013gza}. 

From this brief review, we see that with the recent exception of a preliminary analysis of $\mathcal{N}=1$ theories~\cite{Aharony:2019mbc}, the existing large $N$ analyses have studied Chern-Simons theory coupled to a single boson, a single fermion, or both. There is a gap in the literature concerning large $N$ Chern-Simons theory coupled to multiple flavors. The goal of this work is to start filling this gap. Throughout we take the large $N, k$ limit with the number of fundamentals fixed.

We study three sets of large $N$ theories to leading order in large $N$:
\begin{enumerate}
	\item $U(N)_k$ Chern-Simons theory coupled to $N_f$ fermions and $N_s$ scalars. When $N_f \neq N_s$, we enforce a $SU(N_f)\times SU(N_s)$ global symmetry. This theory is characterized by four approximately marginal couplings. When $N_f = N_s$, we enforce a diagonal $SU(N_f)$ global symmetry, and this theory has five approximately marginal deformations. For specific values of these couplings the theory exhibits $\mathcal{N}=1$ and $\mathcal{N}=2$ supersymmetry.
	\item Setting $N_s$ or $N_f$ to vanish, we consider $U(N)_k$ gauge theory coupled to regular scalars or critical fermions. These theories are characterized by three approximately marginal sextic couplings. 
	\item Finally, from the parent with $N_f$ fermions and $N_s$ scalars we add relevant scalar deformations which drive the model to $U(N)_k$ Chern-Simons theory coupled to $N_f$ regular fermions and $N_s$ Wilson-Fisher scalars.
\end{enumerate}
In each case we compute the thermal free energy as well as the two- and three-point functions of the scalar and fermion bilinears. These are the simplest observables which are not fixed by the symmetries of the problem.\footnote{There are also approximately conserved higher spin operators. In the theory a single fermion or boson, their correlation functions with each other and the scalar/fermion bilinears are fixed by an approximate higher spin symmetry~\cite{Maldacena:2011jn,Maldacena:2012sf}. We expect this to remain true for general $N_f$ and $N_s$.}

The third set of theories is the simplest. To leading order in large $N$ the free energy ``falls apart'' as
\beq
	F = N_f F_{f}(N,k) + N_s F_{s}(N,k)\,,
\eeq
where $F_{f}$ and $F_{s}$ are the thermal free energies obtained in~\cite{Aharony:2012ns,Jain:2013gza} for $U(N)_k$ Chern-Simons theory coupled to a single fermion or a single Wilson-Fisher scalar. There is a similar result for the two- and three-point functions of the fermion and scalar bilinears. These results are consistent with the ``master'' bosonization duality~\eqref{E:master} to leading order in large $N$. 

The second model is less trivial, but we still find evidence for a bosonization duality.

As for the first model, we find that there is a consistent large $N$ duality only when $N_f = N_s$, with interactions that break the $SU(N_f)\times SU(N_f)$ free-field global symmetry to the diagonal $SU(N_f)$. This model may be tuned to the $\mathcal{N}=2$ supersymmetric point, and the duality on the large $N$ conformal manifold may be thought of as arising from a parent Giveon-Kutasov self-duality on Chern-Simons theory coupled to $N_f$ chiral multiplets. Because the $\mathcal{N}=2$ point is believed to be exact at finite $N$, it follows that the induced duality on the rest of the large $N$ conformal manifold will pass all of the finite $N$ tests described above.

We also study the more general dualities of~\cite{Benini:2011mf}, analogous to four-dimensional Seiberg duality. The ``electric'' side is Chern-Simons theory coupled to $N_f$ chiral multiplets in the fundamental representation of the gauge group and $\overline{N}_f$ chiral multiplets in the antifundamental representation. The ``magnetic'' side is Chern-Simons theory coupled to $N_f$ antifundamental multiplets and $\overline{N}_f$ fundamental multiplets, along with $N_f \overline{N}_f$ gauge neutral mesons in the adjoint representation of the $SU(N_f)\times SU(\overline{N}_f)$ global symmetry with a cubic superpotential coupling to the flavor multiplets. Owing to the structure of the matter interactions, which is completely fixed by $\mathcal{N}=2$ supersymmetry, we find the thermal free energy also ``falls apart.'' At leading order in large $N$ the mesons do not contribute to the thermal free energy, which for $U(N)_k$ gauge theory takes the form
\beq
\label{E:N=2free}
	F = N_f F_{\chi}(N,k) + \overline{N}_f F_{\bar{\chi}}(N,k)\,,
\eeq
with $F_{\chi}$ the free energy of $U(N)_k$ gauge theory coupled to a single fundamental chiral multiplet and $F_{\bar{\chi}}$ the free energy of $U(N)_k$ coupled to a single antifundamental chiral multiplet. (In the absence of any other deformations, $F_{\chi} = F_{\bar{\chi}}$, but we find it convenient to separate the free energy this way since we expect the splitting in~\eqref{E:N=2free} to continue to hold in the presence of mass deformations or at finite flavor chemical potential.)

At both finite and zero temperature we focus almost exclusively on the large $N$ fixed points. It would be interesting to further probe these theories by relevant mass and quartic deformations, including to study their Higgs phases as in~\cite{Choudhury:2018iwf}. We have also left $U(N)_k$ Chern-Simons theory coupled to $N_s$ regular scalars and $N_f$ critical fermions for future study. The model with $N_f=N_s$ and diagonal $SU(N_f)$ global symmetry is characterized by $14$ approximately marginal sextic deformations for $N_f >1$.

It would be particularly interesting to study the phase diagram of $U(N)_k$ Chern-Simons coupled to multiple flavors at next-to-leading order (NLO) in the $1/N$ expansion. More generally there are many reasons why it is desirable to go to NLO in $1/N$. One, relevant in the study of Chern-Simons dualities, concerns the master duality in~\eqref{E:master}. That proposal hinges upon the dynamical generation of a fermion mass term in the Higgs phase with a particular sign. This ``extra'' mass is generated only at $O(1/N)$, and its computation would either rule out or give significant evidence for the proposal~\eqref{E:master}.

The remainder of this article is organized as follows. In \underline{Section 2}, we review our conventions for Chern-Simons-matter theory. We compute the planar free energy in \underline{Section 3}, and the two- and three-point functions of scalar/fermion bilinears in \underline{Section 4}. In \underline{Section 5} we show how these results are consistent with duality.

%***********************************************
\section{Preliminaries}
%***********************************************

%***********************************************
\subsection{Chern-Simons and holonomy}
%***********************************************

In this work we consider $U(N)$ Chern-Simons theory at level $k$ coupled to $N_f$ fundamental Dirac fermions and $N_s$ fundamental complex scalars. We work in Euclidean signature, so that the Chern-Simons action is
\beq
	S_{CS} = \frac{i k}{4\pi}\int d^3x \, \epsilon^{\mu\nu\rho}\,\text{tr}_N\left( A_{\mu} \partial_{\nu} A_{\rho} - \frac{2i}{3}A_{\mu}A_{\nu}A_{\rho}\right)\,.
\eeq
We largely follow the conventions of~\cite{Aharony:2012ns}. We take the generators $T^a_{ij}$ of $U(N)$ in the fundamental representation to be anti-Hermitian, with
\beq
	\text{tr}_N(T^a T^b) = -\frac{1}{2}\delta^{ab}\,, \qquad T^a_{ij}T^a_{kl} = - \frac{1}{2}\delta_{il}\delta_{jk}\,,
\eeq
so that the Chern-Simons term is
\beq
	S_{CS} = -\frac{i k}{8\pi}\int d^3x\, \epsilon^{\mu\nu\rho}\left( A_{\mu}^a \partial_{\nu}A_{\rho}^b + \frac{1}{3}f^{abc} A_{\mu}^a A_{\nu}^b A_{\rho}^c\right)
\eeq
In this work we will be interested in the theory at large $N$ and $k$ with the ratio $N/k$ held fixed. The large $N$ counting implies that the observables we consider, like the thermal free energy, of the $U(N)$ Chern-Simons-matter theory coincides with that of $SU(N)$ Chern-Simons theory coupled to the same matter. Similarly, for $U(N) = \big( SU(N) \times U(1)\big)/\mathbb{Z}_N$ Chern-Simons-matter theory, we may consider different levels for the $SU(N)$ and $U(1)$ factors, i.e.
\beq
	U(N)_{k,k+n} = \faktor{\big( SU(N)_k \times U(1)_{kN + N^2n}\big)}{\mathbb{Z}_N}\,.
\eeq
Holding $n$ fixed at large $N$, the same counting implies that the leading order result for many observables is independent of $n$.\footnote{Notable exceptions include the dimension of monopole operators of theories with $U$ gauge group, and of baryon operators in theories with $SU$ gauge group.} Thus in what follows we restrict our attention to $U(N)_k$ Chern-Simons-matter theory.

We regularize the theory (following~\cite{Aharony:2012ns}) by imposing a hard spatial momentum cutoff $\Lambda$, which cures many UV divergences. The remaining divergences, which appear in the sum over modes to obtain the free energy, will be dealt with later. In this regulatory scheme the level $k$ is related to the level $k_{YM}$ that appears when regularizing with a small Yang-Mills term by $k = k_{YM} + N \text{sgn}(k_{YM})$. In the Introduction we used the Yang-Mills convention, but switch here and henceforth. Notably, in the cutoff convention, $|k| \geq N$. Pure Chern-Simons theories famously exhibit level/rank duality~\cite{Naculich:1990pa,Mlawer:1990uv,Nakanishi:1990hj}. In the Yang-Mills regularization, the basic sequence of level/rank dualities reads
\beq
	SU(N)_{-k_{YM}} \quad \leftrightarrow \quad U(k_{YM})_N\,,
\eeq
which in our conventions is
\beq
\label{E:levelrank}
	SU(N)_{-k} \quad \leftrightarrow \quad U(|k|-N)_{k}\,.
\eeq
The conjectured dualities we study between Chern-Simons-matter theories may be thought of as ``flavored'' versions of level/rank duality, in which we begin with a level/rank duality, and then add suitable fundamental matter to both sides. 

The Chern-Simons level obeys a quantization condition. If we define the phase of the fermion functional determinant via the $\eta$-invariant (see e.g.~\cite{Seiberg:2016gmd} for a discussion) so that it is gauge-invariant, then the quantization condition is
\beq
	k \in \mathbb{Z}\,.
\eeq
This level receives a one-loop exact shift after integrating out a massive fermion. In these conventions, integrating out a massive fundamental Dirac fermion of mass $m$ leads to
\beq
	k \to \begin{cases} k & m>0\,, \\ k-1 & m<0\,.\end{cases}
\eeq

When we compute the thermal free energy in the next Section, we place the theory on $\mathbb{R}^2\times \mathbb{S}^1$, using complex coordinates $x^{\pm} = \frac{1}{\sqrt{2}}(x^1 \pm i x^2)$ for space and $x^3\sim x^3 + \beta$ for Euclidean time. In Section~\ref{S:correlators}, when we compute zero temperature correlation functions, we take $\beta\to \infty$. At both finite and zero temperature we gauge-fix $A_- = \frac{A_1+iA_2}{\sqrt{2}}=0$, so that the Chern-Simons term becomes\footnote{In these conventions the invariant tensors are $\delta_{33}=\delta_{+-}=1$ and $\epsilon_{+-3}=i$.}
\beq
\label{E:CSfinal}
	S_{CS} = \frac{k}{4\pi}\int d^3x \, A_+^a \partial_- A_3^a\,.
\eeq

We work in an `t Hooft limit in which $N,k \to \infty$ with $N/k$, $N_f$, and $N_s$ held fixed. The `t Hooft coupling is
\beq
	\lambda = \frac{N}{k}\,,
\eeq
which obeys
\beq
	|\lambda|\leq 1\,,
\eeq
on account of $N \leq |k|$. In terms of $\lambda$ the level/rank duality~\eqref{E:levelrank} between pure Chern-Simons theories maps the `t Hooft coupling of the $SU(N)$ theory to that of the $U(|k|-N)$ theory as
\beq
\label{E:levelrank2}
	|\lambda| \to 1-|\lambda|\,, \qquad \text{sgn}(\lambda) \to - \text{sgn}(\lambda)\,.
\eeq

At zero temperature the gauge field has trivial holonomy in the 3-direction. However, at high temperature, the gauge field acquires a non-trivial holonomy around the thermal circle as first argued by~\cite{Aharony:2012ns}. This holonomy is characterized by the zero mode
\beq
	\mathcal{A}_3 = \frac{1}{V_2}\int_{\mathbb{R}^2} d^2x\, A_3\,,
\eeq
with $V_2$ the regularized volume of $\mathbb{R}^2$. One may use the residual gauge freedom to set $\partial_3\mathcal{A}_3 =0$ and then diagonalize the constant matrix $\mathcal{A}_3$. The holonomy around the thermal circle is then
\beq
	a = i \beta \mathcal{A}_3\,,
\eeq
which is purely real and whose eigenvalues are periodic with periodicity $2\pi$. The end result of~\cite{Aharony:2012ns} is that, at high temperature $\frac{T^2 V_2}{N}\gg 1$ and in the limit with $N\gg 1$ and $\lambda=N/k$ fixed, the eigenvalues of $a$ are uniformly spread around 0 with width $2\pi |\lambda|$, i.e.
\beq
\label{E:holonomy}
	a_{ii} \to a(u) = 2\pi |\lambda| u\,, \qquad u \in \left[ -\frac{1}{2}, \frac{1}{2}\right]\,.
\eeq
Accordingly we replace sums over eigenvalues with an integral over $u$ according to this distribution, 
\beq
\label{E:distribution}
	\sum_{i=1}^N F(a_{ii}) \to N \int_{-\frac{1}{2}}^{\frac{1}{2}} du \, F(2\pi |\lambda| u)\,.
\eeq
Observe that the filled fraction of the circle is $|\lambda|$, so that under level/rank duality the filled fraction is exchanged with its complement. 

%***********************************************
\subsection{Lagrangians}
%***********************************************

In large $N$ Chern-Simons-matter theories there is a distinction between ``regular'' scalar fields and ``critical'' scalar fields. In a theory of ``regular'' scalars $\phi^{\alpha}$ the operator $(\phi^{\dagger}_\alpha \phi^\beta)$ has dimension $\Delta = 1 + O(1/N)$, while in a theory of ``critical' scalars it has dimension $\Delta = 2 + O(1/N)$. As the name suggests, to obtain a theory of regular scalars one simply couples fundamental scalars to a Chern-Simons gauge field. The critical theory is obtained by adding a mass term and quartic interaction and tuning to the non-trivial attractive fixed point. It can be thought of as a gauged version of the Wilson-Fisher theory, and for this reason scalars of this type are sometimes called Wilson-Fisher scalars. There is a similar story for ``regular'' and ``critical'' fermions. 

%***********************************************
\subsubsection{Regular matter}
%***********************************************

Let us begin by considering a Chern-Simons-matter theory with regular fermions and scalars. Denote the fermions as $\psi^m$ and the scalars as $\phi^{\alpha}$, where $m=1,2,..,N_f$ and $\alpha=1,2,..,N_s$. The action of our Chern-Simons matter theories is
\beq
	S = S_{CS} + \int d^3x \Big( \bar{\psi}_m \slashed{D} \psi^m + |D\phi^\alpha|^2  \Big) + S_{int} \,, \qquad D_{\mu} = \partial_{\mu} + A_{\mu}\,,
\eeq
where $S_{CS}$ is given by~\eqref{E:CSfinal} and $S_{int}$ describes interactions amongst the $\psi^m$ and $\phi^\alpha$. The maximal non-abelian global symmetry is $SU(N_f)\times SU(N_s)$. For $N_f\neq N_s$ we consider only those interactions which preserve it. When $N_f = N_s$, with supersymmetric theories in mind, we also allow for interactions which preserve the diagonal global symmetry $SU(N_f) \times SU(N_s)\to SU(N_f)$. At large $N$ the marginal interactions are
\beq
	S_{int} = \int d^3x \left( \frac{1}{3!N^2}V(\phi) + \frac{1}{N}U(\phi,\psi) + \hdots\right)\,,
\eeq
with
\begin{align}
\begin{split}
\label{E:matter}
	V(\phi) & = \lambda_6 (\phi^{\dagger}_\alpha \phi^\alpha)^3 + \lambda_6' (\phi^{\dagger}_\alpha  \phi^\alpha)(\phi^{\dagger}_\beta\phi^\gamma)(\phi^{\dagger}_\gamma \phi^\beta) + \lambda_6'' (\phi^{\dagger}_\alpha \phi^\beta)(\phi^{\dagger}_\beta \phi^\gamma )(\phi^{\dagger}_\gamma \phi^\alpha) \,,
	\\
	U(\phi,\psi) & =\lambda_4 (\bar{\psi}_m  \psi^m)(\phi^{\dagger }_\alpha\phi^\alpha) + \lambda_4' (\bar{\psi}_m \psi^n)(\phi^{\dagger }_n\phi^m) \,,
\end{split}
\end{align}
and the dots indicate terms, like $(\bar{\psi}^m  \phi_\alpha)(\phi^{\dagger \alpha} \psi_m)$, which are marginal at large $N$ but which do not contribute to the large $N$ solution. Parantheses denote the contraction of gauge indices, e.g.
\beq
	(\phi^{\dagger}_{\alpha} \phi^{\beta}) = \phi^{\dagger}_{\alpha i}\phi^{\beta i}\,.
\eeq
 The coupling $\lambda_4'$ only exists in the theory with $N_f = N_s$, and breaks the flavor symmetry down to a single copy of $SU(N_f)$. Clearly when there is a single scalar all of the sextic couplings are identical and, if there is a single fermion, $\lambda_4'$ is also redundant.

When $N_f= N_s$ it is possible to tune to the couplings to preserve $\mathcal{N}=1$ or $\mathcal{N}=2$ supersymmetry. If we preserve $\mathcal{N}=1$ SUSY, then the theory above is $U(N)_k$ Chern-Simons theory coupled to $N_f$ fundamental chiral multiplets $\Phi^m$ with a superpotential
\beq
	W =- \frac{w_1}{2N}\left( \bar{\Phi}_m \Phi^m\right)^2 - \frac{w_2}{2N} \left( \bar{\Phi}_m \Phi^n\right)\left( \bar{\Phi}_n \Phi^m\right)\,,
\eeq
in terms of which the matter couplings in~\eqref{E:matter} are constrained as
\beq
\label{E:supoCouplings}
	\lambda_6 = 6w_1^2\,, \quad \lambda_6' = 12 w_1 w_2\,, \quad \lambda_6'' = 6 w_2^2\,, \quad \lambda_4 =  w_1\,, \quad \lambda_4' = w_2 + 2\pi \lambda\,.
\eeq
$\mathcal{N}=2$ supersymmetry uniquely fixes the superpotential to be
\beq
	W = \frac{2\pi}{k}\left( \bar{\Phi}_m T^a \Phi^m\right)^2 = - \frac{\pi}{k}(\bar{\Phi}_m \Phi^n)(\bar{\Phi}_n\Phi^m)\,,
\eeq
i.e.
\beq
	w_1 = 0\,, \qquad w_2 = 2\pi \lambda\,,
\eeq
so that
\beq
\label{E:N=2couplings}
	\lambda_6 = \lambda_6' = \lambda_4 = 0\,, \qquad \lambda_6'' = 24\pi^2 \lambda^2\,, \qquad \lambda_4' = 4 \pi \lambda\,.
\eeq

There is significant evidence that the $\mathcal{N}=2$ theory above exists and is superconformal. See e.g.~\cite{Gaiotto:2007qi}. There is also significant evidence that it enjoys a self-duality under~\eqref{E:levelrank2}, i.e. under $|\lambda|\to 1-|\lambda|$~\cite{Giveon:2008zn}. We may view the $\mathcal{N}=1$ theory above as a superpotential deformation of the $\mathcal{N}=2$ theory. It has recently been shown~\cite{Aharony:2019mbc} that the manifold of $\mathcal{N}=1$ theories is only approximately superconformal at large $N$, and that an isolated set of points on this manifold correspond to SCFTs. Unlike in four dimensions, the superpotential of a three-dimensional $\mathcal{N}=1$ supersymmetric theory receives perturbative corrections. For a corresponding computation of the beta function of the sextic coupling of Chern-Simons gauge theory coupled to a single regular scalar or a single critical fermion see~\cite{Aharony:2018pjn}.

The $\mathcal{N}=2$ self-duality is the simplest example of a broader set of ``Giveon-Kutasov'' dualities first described in~\cite{Benini:2011mf}. In this broader set the ``electric'' theory is Chern-Simons gauge theory coupled to $N_f$ fundamental chiral multiplets and $\overline{N}_f$ anti-fundamental chiral multiplets, while the ``magnetic'' description is Chern-Simons gauge theory coupled to $\overline{N}_f$ fundamental chiral multiplets, $N_f$ anti-fundamental chiral multiplets, and $N_f \overline{N}_f$ gauge-neutral ``mesons.'' It is easy to also compute the large $N$ thermal free energy of these theories, as we describe in Subsection~\ref{S:N=2}. 

%***********************************************
\subsubsection{Critical matter}
\label{S:criticalmatter}
%***********************************************

We also consider the theory of $N_s$ critical scalars and $N_f$ regular fermions coupled to a Chern-Simons gauge field. At large $N$ this theory is obtained by first setting the couplings in the scalar potential and scalar/fermion interaction to vanish, and adding a mass term for the scalars,
\beq
	S \to S + \int d^3x \,(\sigma_B)^{\alpha}{}_{\beta}( \phi^{\dagger}_{\alpha}\phi^{\beta})\,.
\eeq
We are to compute the exact propagators and thermal free energy as a function of $\sigma^{\alpha}{}_{\beta}$. We then promote $(\sigma_B)^{\alpha}{}_{\beta}$ to be a dynamical field, and extremize the on-shell action with respect to it. The resulting theory is the Legendre transform of the regular theory with respect to the operator $(\phi^{\dagger}_{\alpha} \phi^{\beta})$, which in the deformed theory has dimension $2 + O(1/N)$. In this theory the only marginal operator at large $N$ is
\beq
	(\bar{\psi}_m \phi^{\alpha})(\phi^{\dagger}_{\alpha} \psi^m)\,,
\eeq
which however only contributes at $O(N^0)$ to the propagators and thermal free energy. (If $N_f = N_s$ there is also the operator $(\bar{\psi}_m \phi^m)(\phi^{\dagger}_n \psi^n)$, which is also subleading at large $N$.) Consequently this theory is characterized at leading order in large $N$ by the discrete data $(N,k,N_f, N_s)$.

Similarly we could consider $N_s$ regular scalars and $N_f$ critical fermions, which at large $N$ is the Legendre transform with respect to the operator $(\bar{\psi}_m \psi^n)$. This theory is characterized at large $N$ by 10 marginal couplings in the matter potential when $N_f\neq N_s$, and 14 when $N_f = N_s$. We do not consider it further.

An intermediately complicated theory is that of $N_s$ critical scalars and $N_f$ critical fermions. To obtain it we tune the couplings in the scalar potential and scalar/fermion interaction to vanish, and add mass terms for the scalars and fermions,
\beq
	S \to S + \int d^3x \left(( \sigma_B)^{\alpha}{}_{\beta} (\phi^{\dagger}_{\alpha} \phi^{\beta}) + (\sigma_F)^m{}_n(\bar{\psi}_m \psi^n)\right)\,,
\eeq
along with additional marginal interactions
\beq
\label{E:Ssigma}
	S_{\sigma} =  N \int d^3x \left( \frac{\alpha_3}{3} \text{tr}(\sigma_F)^3 + \frac{\alpha_3'  }{3}\text{tr}(\sigma_F)\text{tr}(\sigma_F^2) + \frac{\alpha_3''}{3} \text{tr}(\sigma_F^3) +\alpha_2\text{tr}(\sigma_B)\text{tr}(\sigma_F) +\alpha'_2\text{tr}(\sigma_B \sigma_F) \right)\,.
\eeq
The last coupling, $\alpha_2'$, only exists when $N_f = N_s$ and we preserve the diagonal $SU(N_f)$ global symmetry. We then promote $\sigma_B$ and $\sigma_F$ to dynamical fields and extremize the on-shell action with respect to them. We expect that this model is dual to the theory of $N_s$ regular fermions and $N_f$ regular scalars with an appropriate action on the five approximately marginal couplings, but we do not investigate it further.

The proposed master duality~\eqref{E:master} of~\cite{Benini:2017aed,Jensen:2017bjo} states that the theory of critical scalars and regular fermions is dual to a theory of critical scalars and regular fermions with, at large $N$,
\beq
	(N,k,N_f,N_s) \quad \leftrightarrow \quad (|k|-N,-k, N_s, N_f)\,.
\eeq
%One might also conjecture a Legendre-transformed version in which a theory of regular scalars and regular fermions is dual to that of critical scalars and critical fermions. Under such a map one expects the parameters $(\lambda_6,\lambda_6',\lambda_6'';\lambda_4,\lambda_4')$ characterizing the matter interactions of the regular theory to get mapped to those $(\alpha_3,\alpha_3',\alpha_3'';\alpha_2,\alpha_2')$ of the critical theory.

In the next Section we will compute the free energy of the theory with regular matter (which for $N_f = N_s$ includes submanifolds with $\mathcal{N}=1$ or $\mathcal{N}=2$ SUSY), critical matter, and the theory with regular fermions and critical scalars.

%***********************************************
\section{Large $N$ free energy}
%***********************************************
%***********************************************
\subsection{Bilocal action}
%***********************************************

The Chern-Simons-matter theories above may be rewritten in terms of a new path integral over bilocal singlet fields, whose equations of motion are the Schwinger-Dyson equations for the exact propagators and whose on-shell value yields the large $N$ free energy. See e.g.~\cite{Jain:2012qi,Jain:2013gza} for nice discussions. In this Subsection we derive those equations of motion and saddle point action using the techniques of~\cite{Jain:2013gza}, which ends up being a minor modification of known results.
% with a single fermion, single boson, or a single fermion and a single boson.

In the ``light-cone'' gauge $A_-=0$ the cubic part of the Chern-Simons term vanishes and so the Chern-Simons-matter action is quadratic in the gauge field. Following~\cite{Giombi:2011kc,Jain:2012qi}, integrating out the gauge field we obtain the action
\begin{align}
\nn
	S = &\int \mathcal{D}^3q \left( \phi^{\dagger}_{\alpha}(-q)(\tilde{q}^2\delta^{\alpha}{}_{\beta}+(\sigma_B)^{\alpha}{}_{\beta}) \phi^{\beta}(q) + \bar{\psi}_m(-q) (i\tilde{q}_{\mu} \gamma^{\mu}\delta^m{}_n+(\sigma_F)^m{}_n)\psi^n(q)\right)
	\\
\nn
	& +N \left\{\int \mathcal{D}^3P\mathcal{D}^3 q_1\mathcal{D}^3q_2 \,C_1(P,q_1,q_2)\chi^m{}_n(P,q_1)\chi^n{}_m(-P,q_2) \right.
	\\
\nn
	& + \int \mathcal{D}^3P_1\mathcal{D}^3P_2\mathcal{D}^3q_1\mathcal{D}^3q_2\mathcal{D}^3q_3 
	\\
\nn
	& \qquad \qquad  \Big[ C_2(P_1,P_2,q_1,q_2,q_3)  \chi^{\alpha}{}_{\beta}(P_1,q_1)\chi^{\beta}{}_{\gamma}(P_2,q_2)\chi^{\gamma}{}_{\alpha}(-P_1-P_2,q_3)
	\\
	& \qquad \qquad \qquad \qquad+ \frac{ \lambda_6 }{6}\chi^{\alpha}{}_{\alpha}(P_1,q_1)\chi^{\beta}{}_{\gamma}(P_2,q_2)\chi^{\gamma}{}_{\beta}(-P_1-P_2,q_3) 
	\\
\nn
	& \qquad \qquad \qquad \qquad \qquad \qquad+ \frac{\lambda_6'}{6} \chi^{\alpha}{}_{\alpha}(P_1,q_1)\chi^{\beta}{}_{\beta}(P_2,q_2)\chi^{\gamma}{}_{\gamma}(-P_1-P_2,q_3)\Big]
	\\
\nn
	& + \int \mathcal{D}^3P\mathcal{D}^3q_1\mathcal{D}^3q_2\, \frac{8\pi i \lambda}{(q_1-q_2)_-}(\xi_-)^m{}_n(P,q_1)(\xi_I)^n{}_m(-P,q_2)
	\\
\nn
	& +\left.2 \int\mathcal{D}^3 P\mathcal{D}^3q_1\mathcal{D}^3q_2\left(   \lambda_4 (\xi_I)^m{}_m(P,q_1) \chi^{\alpha}{}_{\alpha}(-P,q_2) + \lambda_4' (\xi_I)^m{}_n(P,q_1)\chi^n{}_m(-P,q_2)\right)\right\}+ \hdots\,,
\end{align}
where we have defined the singlet bilocal fields
\begin{align}
\begin{split}
	\chi^{\alpha}{}_{\beta}(P,q) &= \frac{1}{N}\phi^{\dagger}_{\beta}\left(\frac{P}{2}-q\right)\phi^{\alpha}\left(\frac{P}{2} + q\right)\,, 
	\\
	(\xi_I)^m{}_n(P,q) & = \frac{1}{2N}\bar{\psi}_n\left(\frac{P}{2}-q\right) \psi^m\left(\frac{P}{2}+q\right)\,,
	\\
	(\xi_-)^m{}_n(P,q) & = \frac{1}{2N}\bar{\psi}_n\left(\frac{P}{2}-q\right) \gamma_- \psi^m\left(\frac{P}{2}+q\right)\,,
\end{split}
\end{align}
and coefficient functions
\begin{align}
\begin{split}
	C_1(P,q_1,q_2) &= 2\pi i \lambda\frac{(-P+q_1+q_2)_3(P+q_1+q_2)_-}{(q_1-q_2)_-}\,,
	\\
	C_2(P_1,P_2,q_1,q_2,q_3) & =4\pi^2\lambda^2 \frac{(P_1-P_2+2q_1+2q_2)_-(P_1+2P_2+2q_2+2q_3)_-}{(P_1+P_2+2q_1-2q_2)_-(P_1-2q_2+2q_3)_-}+ \frac{\lambda_6''}{6}\,,
\end{split}
\end{align}
and the dots correspond to terms which do not contribute to the large $N$ free energy like $(\bar{\psi}_m\phi^{\alpha})(\bar{\phi}_{\alpha}\psi^m)$. The momenta $\tilde{q}$ appearing in the matter kinetic terms are the ``twisted'' momenta that arise on account of the non-trivial holonomy. They are diagonal matrices in color space with entries
\beq
	\tilde{q}_{\mu\, ii} = q_{\mu} - \delta_{\mu}^3 a_{ii}/\beta\,.
\eeq
Further, the Fourier space integration measure $\mathcal{D}^3p$ is that for $\mathbb{R}^2\times \mathbb{S}^1_{\beta}$, 
\beq
	\int \mathcal{D}^3p = \frac{1}{\beta} \int \frac{d^2p}{(2\pi)^2}\sum_{n=-\infty}^{\infty}\,,
\eeq
where the momenta of bosonic fields on the $\mathbb{S}^1$ are quantized as $p_n = \frac{2\pi n}{\beta}$ and fermionic fields as $p_n = \frac{2\pi}{\beta}\left(n+\frac{1}{2}\right)$.

As beautifully explained in~\cite{Jain:2012qi}, one may perform an analogue of a Hubbard-Stratonovich transformation to eliminate the matter fields in favor of new singlet bilocal fields whose dynamics is described by a new path integral. Assuming that the dominant saddle is invariant under translations and rotations, this saddle can be obtained by a somewhat simpler procedure as described in~\cite{Jain:2013gza}. Consider a symmetric saddle for the bilocal fields defined above,
\begin{align}
\begin{split}
	\langle \chi^{\alpha}{}_{\beta}(P,q)\rangle &= (2\pi)^3 \delta^{(3)}(P) \chi^{\alpha}{}_{\beta}(q)\,,
	\\
	 \langle (\xi_I)^m{}_n(P,q)\rangle & = (2\pi)^3 \delta^{(3)}(P) (\xi_{I})^m{}_n(q)\,,
	 \\
	  \langle (\xi_-)^m{}_n(P,q) \rangle &= (2\pi)^3\delta^{(3)}(P) (\xi_{-})^m{}_n(q)\,,
\end{split}
\end{align}
parameterized in terms of the three local functions $\chi(q), \xi_{I}(q),$ and $\xi_{-}(q)$. Plugging such a saddle into the action above leads to an infrared divergent prefactor $\delta^{(3)}(P=0)$ which we interpret as the spacetime volume via $\beta V_2 = (2\pi)^3 \delta^{(3)}(0)$. Note that the term involving $C_1$ vanishes due to symmetry. Next we introduce new fields $((G_B)^{\alpha}{}_{\beta},(\Sigma_B)^{\alpha}{}_{\beta})$ and $2\times 2$ matrix valued fields $((G_{F})^{\alpha}{}_{\beta},(\Sigma_F)^{\alpha}{}_{\beta})$ to decouple the interactions. Here
\beq
	G_F = G_{I}I_2 + G_{-}\gamma^-\,, \qquad \Sigma_F = \Sigma_I I_2 + \Sigma_+ \gamma^+\,.
\eeq
The new action is
\begin{align}
\nn
	\tilde{S} = &  \int \mathcal{D}^3q \left( \phi^{\dagger}_{\alpha}(-q)(\tilde{q}^2 \delta^{\alpha}{}_{\beta}+ (\tilde{\Sigma}_B)^{\alpha}{}_{\beta}(q)) \phi^{\beta}(q) + \bar{\psi}_m(-q)(i \tilde{q}_{\mu}\gamma^{\mu}\delta^m{}_n  + (\tilde{\Sigma}_F)^m{}_n)\psi^n(q)\right)
	\\
\nn
	& +N \beta V_2\left\{ \int  \mathcal{D}^3q_1\mathcal{D}^3q_2\mathcal{D}^3q_3  \Big[ C_2(q_1,q_2,q_3) (G_B)^{\alpha}{}_{\beta}(q_1)(G_B)^{\beta}{}_{\gamma}(q_2)(G_B)^{\gamma}{}_{\alpha}(q_3)\right.
	\\
	\nn
	& \qquad \qquad + \frac{\lambda_6}{6}(G_B)^{\alpha}{}_{\alpha}(q_1) (G_B)^{\beta}{}_{\beta}(q_2) (G_B)^{\gamma}{}_{\gamma}(q_3)
	\\
	& \qquad \qquad \qquad \qquad+ \frac{\lambda_6'}{6}(G_B)^{\alpha}{}_{\alpha}(q_1)(G_B)^{\beta}{}_{\gamma}(q_2)(G_B)^{\gamma}{}_{\beta}(q_3)\Big]
	\\
	\nn
	& \qquad\qquad + \int \mathcal{D}^3q_1 \mathcal{D}^3q_2 \frac{8\pi i \lambda}{(q_1-q_2)_-}(G_{-})^m{}_n(q_1)(G_{I})^n{}_m(q_2)
	\\
\nn
	&\qquad \qquad+ 2\int \mathcal{D}^3q_1 \mathcal{D}^3 q_2\left( \lambda_4 (G_B)^{\alpha}{}_{\alpha}(q_1)(G_I)^m{}_m(q_2)  + \lambda_4' (G_B)^m{}_n(q_1)(G_{I})^n{}_m(q_2)\right) 
	\\
\nn
	& \qquad\qquad-\left.  \int \mathcal{D}^3q \,\text{tr}\Big( 	\Sigma_B(q) G_B(q)+\Sigma_F(q)G_F(q)\Big) \right\}+  \hdots\,,
\end{align}
with
\beq
	\tilde{\Sigma}_B = \Sigma_B + \sigma_B\,, \qquad \tilde{\Sigma}_F = \Sigma_F + \sigma_F I_2\,,
\eeq
and
\beq
	C_2(q_1,q_2,q_3) = C_2(0,0,q_1,q_2,q_3) =4\pi^2 \lambda^2 \frac{(q_1+q_2)_-(q_2+q_3)_-}{(q_1-q_2)_-(q_3-q_2)_-} + \frac{\lambda_6''}{6}\,.
\eeq
In the last line, the trace is performed over flavor and spinor indices. To see the equivalence with the original action integrate out $\Sigma_B$ and $\Sigma_F$. Doing so sets $G_B(q) = \chi(0,q)$ and $G_F(q) = \xi_I(0,q)I_2 + \xi_-(0,q)\gamma^-$. Thus we see that the saddle-point solutions for $G_B$ and $G_F$ are the color singlet parts of the exact large $N$ bosonic and fermionic propagators.

The new action is quadratic in the bosons and fermions and so we integrate them out, giving
\begin{align}
\nn
	\tilde{S} = & \beta V_2 \text{tr}_N \int \mathcal{D}^3q \,\text{tr}\Big( \ln (\tilde{q}^2+\tilde{\Sigma}_B(q)) -\ln (i \tilde{q}_{\mu}\gamma^{\mu} + \tilde{\Sigma}_F(q))\Big)
	\\
\nn
	& + N \beta V_2 \left\{ \int \mathcal{D}^3q_1\mathcal{D}^3q_2\mathcal{D}^3q_3 \Big[ C_2(q_1,q_2,q_3)(G_B)^{\alpha}{}_{\beta}(q_1)(G_B)^{\beta}{}_{\gamma}(q_2)(G_B)^{\gamma}{}_{\alpha}(q_3)\right.
	\\
\nn
	& \qquad \qquad\qquad + \frac{\lambda_6}{6}(G_B)^{\alpha}{}_{\alpha}(q_1)(G_B)^{\beta}{}_{\beta}(q_2)(G_B)^{\gamma}{}_{\gamma}(q_3)
	\\
	\label{E:bilocalS}
	&\qquad \qquad \qquad\qquad \qquad + \frac{\lambda_6'}{6}(G_B)^{\alpha}{}_{\alpha}(q_1)(G_B)^{\beta}{}_{\gamma}(q_2)(G_B)^{\gamma}{}_{\beta}(q_3)\Big]
	\\
\nn
	& \qquad+ \int \mathcal{D}^3q_1 \mathcal{D}^3q_2\,  \frac{8\pi i \lambda}{(q_1-q_2)_-}(G_{-})^m{}_n(q_1)(G_{I})^n{}_m(q_2) 
	\\
\nn 
	& \qquad  + 2\int \mathcal{D}^3q_1\mathcal{D}^3q_2\Big(  \lambda_4(G_B)^{\alpha}{}_{\alpha}(q_1)(G_{I})^m{}_m(q_2)+\lambda_4'(G_B)^m{}_n(q_1)(G_I)^n{}_m(q_2)\Big)
	\\
\nn
	& \qquad-\left. \int \mathcal{D}^3q \,\text{tr}\Big( \Sigma_B(q)G_B(q) + \Sigma_F(q)G_F(q)\Big)\right\} + \hdots\,.
\end{align}
The traces in the first and last lines are performed over flavor and spinor indices.

At this stage, for generic bosonic masses $(\sigma_B)^{\alpha}{}_{\beta}$ and fermionic masses $(\sigma_F)^m{}_n$, the $SU(N_f)\times SU(N_s)$ flavor symmetry is broken to $U(1)^{N_f + N_s-2}$. However, recall that we are ultimately interested in (i.) the $SU(N_f)\times SU(N_s)$ symmetric theories of regular matter, (ii.) the theory with either regular scalars or critical fermions, or (iii.) the theory with regular fermions and critical scalars. 

In the first case, the fixed point is at $\sigma_B = \sigma_F = 0$. In the second, the massless theory with regular scalars alone has $\sigma_B = 0$; the theory with critical fermions is obtained from that of regular massive fermions with no matter potential. Setting the $\lambda$'s to vanish and diagonalizing the fermion mass $\sigma_F$, we see from the action~\eqref{E:bilocalS} that the $N_f$ fermions only couple to themselves and not each other. Finally, the theory of critical scalars and regular fermions is obtained from the theory with regular matter and no matter potential along with a boson mass $\sigma_B$. Diagonalizing it and setting the $\lambda$'s to vanish, we also see that the individual scalars and fermions are effectively self-interacting, and do not couple to the other flavors at leading order in large $N$.

In each case this simplification allows us to easily solve the equations of motion and on-shell action.

Let us deal with each of these theories in turn. We begin with the theory of regular massless matter, for which we make a $U(N_f)\times U(N_s)$ ansatz for the bilocal fields:
\beq
	(G_B(q))^{\alpha}{}_{\beta} = G_B(q) \delta^{\alpha}{}_{\beta}\,, \qquad (\Sigma_B(q))^{\alpha}{}_{\beta} = \Sigma_B(q) \delta^{\alpha}{}_{\beta}\,, \qquad (\sigma_B)^{\alpha}{}_{\beta} =0
	% \sigma_B \delta^{\alpha}{}_{\beta}\,,
\eeq
and similarly for the fermionic objects. We will tackle the other cases later in the Section.%Assuming that the dominant saddle is $U(N_f)\times U(N_s)$-symmetric, the extremization over the matrix-valued fields $(\sigma_B)^{\alpha}{}_{\beta}$ and $(\sigma_F)^m{}_n$ amounts to extremization over the scalar parameters $\sigma_B$ and $\sigma_F$.

With this substitution at hand the effective action simplifies further, becoming
\begin{align}
	\label{E:bilocalS2}
	\tilde{S} = & \beta V_2 \text{tr}_N \int \mathcal{D}^3q \Big(N_s \ln (\tilde{q}^2+%\tilde{\Sigma}_B(q)
	\Sigma_B(q)) -N_f\text{tr}\ln (i \tilde{q}_{\mu}\gamma^{\mu} +% \tilde{\Sigma}_F(q)
	\Sigma_F(q))\Big)
	\\
	\nn
	& + N \beta V_2 \left\{ N_s\int \mathcal{D}^3q_1\mathcal{D}^3q_2\mathcal{D}^3q_3 \left(\frac{N_s^2\lambda_6+N_s\lambda_6'}{6}+ C_2(q_1,q_2,q_3)\right)G_B(q_1)G_B(q_2)G_B(q_3)\right.
	\\
	\nn
	& \qquad+N_f \int \mathcal{D}^3q_1 \mathcal{D}^3q_2\,  \frac{8\pi i \lambda }{(q_1-q_2)_-}G_{-}(q_1)G_{I}(q_2) + 2\Big(   N_s\lambda_4+\lambda_4'\Big)G_B(q_1)G_{I}(q_2)
	\\
	\nn
	& \qquad-\left. \int \mathcal{D}^3q \Big(N_s \Sigma_B(q)G_B(q) + N_f\text{tr}_{\rm f}(\Sigma_F(q)G_F(q))\Big)\right\} + \hdots\,.
\end{align}
The trace in the last line is over spinor indices alone.

%***********************************************
\subsubsection{Equations of motion}
%***********************************************

The equations of motion for $\Sigma_B$ and $\Sigma_F$ are
\beq
\label{E:eom1}
	G_B(q) =\frac{\text{tr}_N}{N}\left( \frac{1}{\tilde{q}^2 +
	% \tilde{\Sigma}_B(q)
	\Sigma_B(q)}\right)\,, \qquad G_F(q)=-\frac{\text{tr}_N}{N}\left(\frac{1}{i \tilde{q}_{\mu}\gamma^{\mu} + %\tilde{\Sigma}_F(q)
	\Sigma_F(q)}\right)\,,
\eeq
which upon using the piecewise-uniform eigenvalue distribution~\eqref{E:distribution} is
\beq
	G_B(q) = \int_{-\frac{1}{2}}^{\frac{1}{2}}du \frac{1}{\tilde{q}^2+%\tilde{\Sigma}_B(q)
	\Sigma_B(q)}\,, \qquad G_F(q) = -\int_{-\frac{1}{2}}^{\frac{1}{2}}du\frac{1}{i\tilde{q}_{\mu}\gamma^{\mu} +%\tilde{\Sigma}_F(q)
	\Sigma_F(q)}\,.
\eeq
The equations of motion for $G_B$ and $G_F$ give
\begin{align}
\nn
	\Sigma_B(q) &= \int\mathcal{D}^3p_1 \mathcal{D}^3p_2\left( C_2(q,p_1,p_2)+C_2(p_1,q,p_2)+C_2(p_1,p_2,q)+\frac{N_s^2\lambda_6+N_s\lambda_6'}{2}\right) 
	\\
	\nn
	& \qquad \qquad \qquad \qquad \times G_B(p_1)G_B(p_2) + 2(N_f\lambda_4+\lambda_4') \int \mathcal{D}^3p\, G_I(p)\,,
	\\
	\label{E:eom2}
	\Sigma_I(q) & = -4\pi i\lambda \int \mathcal{D}^3p \frac{G_-(p)}{(q-p)_-} + (N_s\lambda_4+\lambda_4')\int \mathcal{D}^3p\,G_B(p)\,,
	\\
	\nn
	\Sigma_+(q) & = 4\pi i \lambda \int\mathcal{D}^3p\frac{G_I(p)}{(q-p)_-}\,.
\end{align}

Plugging in the expressions~\eqref{E:eom1} for $G_B$ and $G_F$ we can express $\Sigma_B$ as
\beq
	\Sigma_B(q) = (\text{a})+(\text{b})+(\text{c})+(\text{d})+(\text{e})\,,
\eeq
with
\begin{align}
\nn
	(\text{a}) & = \frac{4\pi^2}{k^2}\left[\int\mathcal{D}^3p \frac{(p+q)_-}{(p-q)_-}\text{tr}_N\left(\frac{1}{\tilde{p}^2+%\tilde{\Sigma}_B(p)
	\Sigma_B(p)}\right)\right]^2\,,
	\\
\nn
	(\text{b}) = (\text{c}) & = \frac{4\pi^2}{k^2}\int \mathcal{D}^3p_1 \mathcal{D}^3p_2 \frac{(q+p_1)_-(p_1+p_2)_-}{(q-p_1)_-(p_2-p_1)_-}\text{tr}_N\left(\frac{1}{\tilde{p}_1^2+
	%\tilde{\Sigma}_B(p_1)
	\Sigma_B(p_1)}\right)\text{tr}_N\left(\frac{1}{\tilde{p}_2^2+%\tilde{\Sigma}_B(p_2)
	\Sigma_B(p_2)}\right) \,,
	\\
	\label{E:bosonicSigma}
	(\text{d}) & = \frac{\lambda_6 N_s^2+\lambda_6'N_s+\lambda_6''}{2}\left[ \frac{1}{N}\int\mathcal{D}^3p \,\text{tr}_N\left(\frac{1}{\tilde{p}^2+%\tilde{\Sigma}_B(p)
	\Sigma_B(p)}\right)\right]^2\,,
	\\
\nn
	(\text{e}) & = -\frac{N_f\lambda_4+\lambda_4'}{N}\int \mathcal{D}^3p\, \text{tr}\left(\frac{1}{i\tilde{p}_{\mu}\gamma^{\mu}+%\tilde{\Sigma}_F(p)
	\Sigma_F(p)}\right)\,,
\end{align}
where in the last line the trace is taken over color and spinor indices. Similarly we have
\begin{align}
\begin{split}
\label{E:fermionSigma}
	\Sigma_+(q) & = \frac{2\pi i }{k} \int \mathcal{D}^3p\frac{1}{(q-p)_-}\text{tr}\left(\frac{1}{i\tilde{p}_{\mu}\gamma^{\mu}+%\tilde{\Sigma}_F
	\Sigma_F}\right) = \frac{4\pi i}{k}\int \mathcal{D}^3p\frac{1}{(q-p)_-}\text{tr}_N\left( \frac{%\tilde{\Sigma}_I
	\Sigma_I}{(\tilde{p}-i \Sigma_F)^2 + %\tilde{\Sigma}_I^2
	\Sigma_I^2}\right)\,,
	\\
	\Sigma_I(q) & = -\frac{2\pi i}{k}\int\mathcal{D}^3p \frac{1}{(q-p)_-}\text{tr}\left( \gamma^+\frac{1}{i\tilde{p}_{\mu}\gamma^{\mu}+%\tilde{\Sigma}_F
	\Sigma_F}\right) + \frac{N_s\lambda_4+\lambda_4'}{N}\int\mathcal{D}^3p \,\text{tr}_N\left( \frac{1}{\tilde{p}^2+%\tilde{\Sigma}_B
	\Sigma_B}\right)
	\\
	& = \frac{4\pi i}{k }\int \mathcal{D}^3p \frac{1}{(q-p)_-}\text{tr}_N\left( \frac{ip_-}{(\tilde{p}-i \Sigma_F)^2+%\tilde{\Sigma}_I^2
	\Sigma_I^2}\right) + \frac{N_s\lambda_4+\lambda_4'}{N}\int\mathcal{D}^3p \,\text{tr}_N\left(\frac{1}{\tilde{p}^2+%\tilde{\Sigma}_B
	\Sigma_B}\right)\,.
\end{split}
\end{align}
Comparing these equations with the ``bootstrap equations'' for the bosonic and fermionic self-energies obtained in~\cite{Aharony:2012ns}, we readily see that (i.) our $\Sigma_B$ and $\Sigma_F$ may be interpreted as the bosonic and fermionic self-energies of the exact large $N$ propagators via
\begin{align}
\begin{split}
	\langle \phi^{\alpha}(q)\phi^{\dagger}_{\beta}(p)\rangle& = (2\pi)^3 \delta^{(3)}(p+q) \delta^{\alpha}{}_{\beta} \text{tr}_N\left(\frac{1}{\tilde{q}^2 + 
	%\tilde{\Sigma}_B(q)
	\Sigma_B(q)}\right)\,,
	\\
	\langle \psi^m(q)\bar{\psi}_n(p)\rangle & = (2\pi)^3 \delta^{(3)}(p+q)\delta^m{}_n \text{tr}_N\left( \frac{1}{i\tilde{q}_{\mu}\gamma^{\mu} + %\tilde{\Sigma}_F(q)
	\Sigma_F(q)}\right)\,,
\end{split}
\end{align}
and (ii.) these are the same equations they find (for a single fundamental boson and fermion) with a flipped sign of the self-energies,
\beq
	\Sigma_{B\,\rm here} = -\Sigma_{B\, \rm there}\,, \qquad \Sigma_{F\, \rm here} = -\Sigma_{F\,\rm there}\,, 
\eeq
along with the replacements
\beq
\label{E:replace}
	\lambda_{6\,\rm there} \to N_s^2\lambda_6+N_s \lambda_6'+\lambda_6''\,, 
\eeq
and, in the scalar equation, $\lambda_{4\, \rm there} \to N_f \lambda_4 + \lambda_4'$, while in the fermionic equation $\lambda_{4\,\rm there} \to N_s \lambda_4 + \lambda_4'$. These replacements effectively account for the multiple flavors. Indeed, rather than obtaining the equations of motion for $(G_B,\Sigma_B)$ and $(G_F,\Sigma_F)$ we could simply have summed up the planar diagrams with the same result. 

%***********************************************
\subsubsection{On-shell action}
%***********************************************

Using the equations of motion, we can bring the on-shell action~\eqref{E:bilocalS} into the form
\beq
\label{E:onShellS}
	\tilde{S} =S_B + S_F + S_{BF}\,,
\eeq
where
\begin{align}
\nn
	S_B &= N N_sV_2 \int_{-\frac{1}{2}}^{\frac{1}{2}}du \int \frac{d^2q}{(2\pi)^2}\sum_{n=-\infty}^{\infty} \left( \ln (\tilde{q}^2+\tilde{\Sigma}_B) - \frac{2}{3}\frac{\Sigma_B}{\tilde{q}^2+
	%\tilde{\Sigma}_B
	\Sigma_B}\right)\,,
	\\
	\label{E:onshell}
	S_F & = NN_fV_2\int_{-\frac{1}{2}}^{\frac{1}{2}} du \int \frac{d^2q}{(2\pi)^2}\sum_{n=-\infty}^{\infty} \text{tr}_{\rm f} \left(- \ln (i\tilde{q}_{\mu}\gamma^{\mu} +\tilde{\Sigma}_F) +\frac{1}{2}\Sigma_F\frac{1}{i\tilde{q}_{\mu}\gamma^{\mu}+
	%\tilde{\Sigma}_F
	\Sigma_F}\right)\,,
	\\
	\nn
	S_{BF} & = -\frac{\beta NN_fV_2(N_s\lambda_4+ \lambda_4')}{6}\left( \int_{-\frac{1}{2}}^{\frac{1}{2}}du \int \mathcal{D}^3q \,\frac{1}{\tilde{q}^2+	%\tilde{\Sigma}_B
	\Sigma_B}\right)\left(\int_{-\frac{1}{2}}^{\frac{1}{2}}du\int\mathcal{D}^3p \,\text{tr}_{\rm f} \left( \frac{1}{i\tilde{p}_{\mu}\gamma^{\mu}+%\tilde{\Sigma}_F
	\Sigma_F}\right)\right)\,.
\end{align}
For the theory of critical scalars coupled to a Chern-Simons gauge field, we must set $\lambda_6 = \lambda_6'=\lambda_6''=\lambda_4=\lambda_4' = 0$.

%***********************************************
\subsection{Exact propagators and free energy}
\label{S:freeenergy}
%***********************************************

We now proceed to solve the equations of motion~\eqref{E:bosonicSigma} and~\eqref{E:fermionSigma} for the self-energies $\Sigma_B$ and $\Sigma_F$. These equations are the same as those encountered in~\cite{Aharony:2012ns} with suitable replacements of parameters. We will mostly adapt their solution in what follows.

Taking the derivative of the equation of motion~\eqref{E:bosonicSigma} for $\Sigma_B$ with respect to $q$ and using
\beq
	\frac{\partial}{\partial q_+}\frac{1}{q_-} = 2\pi \delta^{(2)}(q)\,, \qquad \frac{\partial q_s}{\partial q_+} = \frac{q_-}{q_s}\,,
\eeq
one finds that $\Sigma_B$ is independent of $q$. Similar manipulations of~\eqref{E:fermionSigma} for $\Sigma_F$ reveal that\footnote{Our conventions for $f$ and $g$ below are related to those of~\cite{Aharony:2012ns} by $f_{\rm us} = -f_{\rm them}$ and $g_{\rm us} = -g_{\rm them}$.}
\begin{align}
\begin{split}
	%\tilde{\Sigma}_B(q) 
	\Sigma_B(q)&= \beta^{-2} \mu_B^2\,,
	\\
	%\tilde{\Sigma}_I(q) 
	\Sigma_I(q)& = f(y) q_s\,,
	\\
	\Sigma_+(q) & = ig(y) q_+\,,
	\\
	g + f^2 &= \frac{\mu_F^2}{y^2}\,,
\end{split}
\end{align}
with $y = \beta q_s$ and where $\mu_B$, $\mu_F$, $f$, and $g$, are all functions of the couplings and the matter content. The quantities $\mu_B$ and $\mu_F$ may be interpreted as thermal pole masses. In equation~\eqref{E:bosonicSigma} we may then substitute $q=0$, for which $\text{(b)} =\text{(c)}=0$. The basic integrals remaining in that equation of motion are
\begin{align}
\label{E:basicIntegrals}
	I_B = \frac{1}{N}\int \mathcal{D}^3p \,\text{tr}_N\left( \frac{1}{\tilde{p}^2+%\tilde{\Sigma}_B
	\Sigma_B}\right) &= -\frac{h_B}{4\pi\beta| \lambda|}\,,
	\\
	\nn
	I_F = \frac{1}{N}\int \mathcal{D}^3p \,\text{tr}\left( \frac{1}{i\tilde{p}_{\mu}\gamma^{\mu}+%\tilde{\Sigma}_F
	\Sigma_F}\right) & = - \frac{1}{4\pi \beta^2\lambda}\left( h_F +2\mathcal{H}_F\right)h_F \,,
	%\KJ{+\frac{\lambda \mu_F^2}{4\pi \beta^2}?}\,,
\end{align}
where
\begin{align}
\begin{split}
\label{E:hBhF}
	h_B &= |\lambda |\mu_B + \frac{1}{\pi i}\left( \text{Li}_2(e^{-\mu_B-\pi i |\lambda|}) - \text{c.c.}\right)\,,
	\\
	h_F & =| \lambda| \mu_F + \frac{1}{\pi i} \left( \text{Li}_2(-e^{-\mu_F - \pi i| \lambda|}) - \text{c.c.}\right)\,,
	\\
	\mathcal{H}_F &= %\beta \sigma_F 
	- \frac{N_s\lambda_4+\lambda_4'}{4\pi \lambda}h_B\,,
\end{split}
\end{align}
and we have chosen a convention for which $\mu_B$ and $\mu_F$ are both positive. Further, the function $f$ is given by
\beq
	y f(y) = \lambda \sqrt{y^2+\mu_F^2} +\frac{1}{\pi i}\left[ \text{Li}_2(-e^{-\sqrt{y^2+\mu_F^2}-\pi i \lambda}) - \text{c.c.}\right] +\text{sgn}(\lambda)\mathcal{H}_F\,,
\eeq
which comes from evaluating the last line of~\eqref{E:fermionSigma}. In evaluating the integrals~\eqref{E:basicIntegrals} we integrate up to a hard cutoff for spatial momenta and cancel divergences with mass counterterms. To evaluate the second integral, it is convenient to first sum over Matsubara modes, then integrate over $u$, giving
\beq
	I_F = \frac{i}{2\pi^2\beta^2 \lambda}\int_0^{\beta\Lambda} dx\, x f(x)\, \frac{x}{\sqrt{x^2+\mu_F^2}}\left[ \ln \cosh\left( \frac{\sqrt{x^2+\mu_F^2}-\pi i \lambda}{2}\right) - \text{c.c.}\right]\,,
\eeq
with $x = \beta p_s$, and then use that
\beq
	\pi i \frac{\partial (xf(x))}{\partial x} =- \frac{x}{\sqrt{x^2+\mu_F^2}}\left[ \ln \cosh\left( \frac{\sqrt{x^2+\mu_F^2}-\pi i \lambda}{2}\right)- \text{c.c.}\right]\,.
\eeq
The equations of motion then become transcendental equations for $\mu_B$ and $\mu_F$,
\begin{align}
\begin{split}
\label{E:finalFormEOMs}
	\mu_B^2% - \beta^2\sigma_B 
	& = \left( \frac{\hat{\lambda}}{2\lambda}h_B\right)^2 + \frac{N_f\lambda_4+\lambda_4'}{4\pi \lambda}\left( h_F +2\mathcal{H}_F\right)h_F\,,
	\\
	\mu_F  & = h_F +\mathcal{H}_F\,,
\end{split}
\end{align}
where we have defined
\beq
	\hat{\lambda}^2 =\lambda^2 + \frac{N_s^2 \lambda_6 + N_s \lambda_6' + \lambda_6''}{8\pi^2}\,,
\eeq
and we choose the sign for $\mu_F$ so that it is positive.

Having solved the equations of motion we presently evaluate the on-shell action, which gives us the thermal free energy $\beta F$ to leading order at large $N$. We separately evaluate the terms in the action $\tilde{S}$ in Eq.~\eqref{E:onShellS}. The one-loop determinant terms in $S_B$ and $S_F$ are simply $N_s$ and $N_f$ times the corresponding terms in the computation of~\cite{Aharony:2012ns}. As for the rest of the bosonic term $S_B$, it is given by
\beq
	-\frac{2N_s V_2\beta}{3}\int \mathcal{D}^3q\, \text{tr}_N \left( \frac{\Sigma_B}{\tilde{q}^2 + %\tilde{\Sigma}_B
	\Sigma_B}\right) =- \frac{2NN_s V_2}{3\beta}\mu_B^2
	%(\mu_B^2 - \tilde{\sigma}_B)
	 I_B = \frac{N N_s V_2}{6\pi \beta^2 |\lambda|}h_B%(\mu_B^2 - \tilde{\sigma}_B)
	\mu_B^2\,.
\eeq
%where $\Sigma_B = \beta^{-2}\mu_B^2 - \sigma_B$ and $\tilde{\sigma}_B = \beta^{-2}\sigma_B$. T
Thus
\begin{align}
\begin{split}
	S_B = -\frac{N N_s V_2}{2\pi \beta^2}&\left\{% \frac{1}{3}\left( \mu_B^3 - \frac{h_B(\mu_B^2-\tilde{\sigma}_B)}{\lambda}\right) 
	\frac{\mu_B^2}{3}\left( \mu_B -\frac{h_B}{\lambda}\right)+ \frac{1}{\pi i \lambda}\int_{\mu_B}^{\infty} dy\,y\left[ \text{Li}_2\left( e^{-y+\pi i \lambda}\right) - \text{c.c.}\right]\right\}
\end{split}
\end{align}
Similarly, after some manipulations we find that the fermionic contribution to the on-shell action is
\begin{align}
\begin{split}
	S_F = \frac{N N_f V_2}{2\pi \beta^2}&\left\{ \frac{\mu_F^3}{3}\left( 1\mp \frac{1}{\lambda} \right)  +\frac{1}{4\lambda} \left( \mu_F^2\mathcal{H}_F+%\mu_F^2\tilde{\sigma}_F) 
	\frac{\mathcal{H}_F^3}{3} % \tilde{\sigma}_F \mathcal{H}_F^2
	\right) \right.
	\\
	& \qquad + \left.\frac{1}{\pi i \lambda} \int_{\mu_F}^{\infty} dy \, y \left[ \text{Li}_2\left( -e^{-y + \pi i \lambda} \right) - \text{c.c.}\right]\right\}\,.
\end{split}
\end{align}
%where $\tilde{\sigma}_F = \beta^{-1} \sigma_F$. 
Finally, the interaction term $S_{BF}$ is
\beq
	S_{BF} = - \frac{N N_f(N_s \lambda_4+ \lambda_4')V_2}{6} I_B I_F= -\frac{N N_f (N_s\lambda_4+\lambda_4')V_2}{96\pi^2\beta^2\lambda^2} h_Bh_F(h_F+2\mathcal{H}_F)\,.
\eeq

%For now let us set the bare masses $\sigma_B$ and $\sigma_F$ to vanish. 
Putting the pieces together, and after extensive use of Eqs.~\eqref{E:hBhF} and~\eqref{E:finalFormEOMs}, we arrive at the large $N$ free energy for the theory of regular matter. It is 
\begin{align}
\nn
	\beta F = & -\frac{N V_2}{2\pi^2 i \beta^2|\lambda|} \left\{  \frac{N_s \mu_B^2}{3}\left[ \text{Li}_2\left( e^{-\mu_B + \pi i| \lambda|}\right) - \text{c.c.}\right] - \frac{N_f\mu_F^2}{3}\left[ \text{Li}_2\left( -e^{-\mu_F + \pi i |\lambda|}\right) - \text{c.c.}\right]\right.
	\\
	\label{E:Fregreg}
	& \left. + N_s \int_{\mu_B}^{\infty}dy\,y \left[ \text{Li}_2\left(e^{-y+\pi i |\lambda|}\right) - \text{c.c.}\right]- N_f \int_{\mu_F}^{\infty}dy\,y \left[ \text{Li}_2\left(-e^{-y+\pi i |\lambda|}\right) - \text{c.c.}\right] \right\}\,.
\end{align}
All information about the self-interactions $\lambda_6, \lambda_6'$, etc., is subsumed into the values $\mu_B$ and $\mu_F$ of the thermal pole masses, which are determined by~\eqref{E:finalFormEOMs}. Since $N/|\lambda| = |k|$, this presentation is tailor-made for a Chern-Simons duality under 
\beq
	N \to |k|-N\,, \qquad k\to -k\,, \qquad N_s\leftrightarrow N_f\,, \qquad \mu_B \leftrightarrow \mu_F\,,
\eeq
However, as we discussed in the Introduction, the spectrum of gauge-invariant operators of a theory with $N_f$ regular fermions and $N_s$ regular scalars can match those of a dual with $N_f'$ fermions and $N_s'$ scalars only if $N_s = N_s'$ and $N_f = N_f'$. Putting this together with our result for the free energy, we see that this duality is only tenable if there are equal numbers of fermions and scalars, $N_f = N_s$. In other words, if it is a self-duality. We discuss this further in Section~\ref{S:testRR}.

%***********************************************
\subsection{Regular scalars or critical fermions}
\label{S:critferm}
%***********************************************

The free energy of the theory with $N_f$ regular scalars is given by~\eqref{E:Fregreg} upon setting the number of fermions $N_f = 0$.

To reach the theory of $N_f$ critical fermions, we first set $N_s = 0$ along the couplings of the matter potential, and allow for a bare fermion mass $(\sigma_F)^m{}_n$. In terms of the eigenvalues of the bare mass matrix, which we label as $\sigma_m$, it is easy to repeat the analysis above for the theory of regular matter. The resulting free energy is the sum of free energies for the individual fermions, and reads
\begin{align}
\begin{split}
	\beta F = \frac{NV_2}{2\pi \beta^2}& \left\{ \sum_{m=1}^{N_f}\left(\frac{\mu_m^2 \tilde{\sigma}_m}{6\lambda} - \frac{\tilde{\sigma}_m^3}{6\lambda}+ \frac{\mu_m^2}{3\pi i |\lambda|} \left[ \text{Li}_2\left( - e^{-\mu_m + \pi i |\lambda|}\right)-  \text{c.c.}\right]   \right.\right.
	\\
	&\left.\left.  \qquad \qquad \qquad \qquad+  \frac{1}{\pi i|\lambda|}\int_{\mu_m}^{\infty} dy \,y\left[ \text{Li}_2\left( -e^{-y+\pi i |\lambda|}\right)-\text{c.c.}\right] \right)\right\}\,,
\end{split}
\end{align}
where we have labeled the thermal pole mass of the $m^{\rm th}$ fermion as $\mu_m$ and $\tilde{\sigma}_m = \beta \sigma_m$. The thermal pole masses satisfy
\beq
\label{E:critfermpolemass}
	\pm \mu_m = \tilde{\sigma}_m +| \lambda| \mu_m -\frac{1}{\pi i}\left[ \text{Li}_2 \left( -e^{-\mu_m + \pi i |\lambda|}\right) - \text{c.c.}\right]\,.
\eeq
Following the discussion in Subsection~\ref{S:criticalmatter}, we obtain the critical theory by adding the triple-trace interactions $\Delta \mathcal{L} = \frac{N}{3!}\left( \lambda_{6F}\text{tr}(\sigma_F)^3 + \lambda_{6F}'\text{tr}(\sigma_F)\text{tr}(\sigma_F^2)+\lambda_{6F}''\text{tr}(\sigma_F^3)\right)$, which deforms the free energy by
\beq
	\Delta (\beta F) = \frac{NV_2}{3!\beta^2}\left(\lambda_{6F} \left( \sum_{m=1}^{N_f}\tilde{\sigma}_m\right)^3+\lambda_{6F}'\left(\sum_{m=1}^{N_f}\tilde{\sigma}_m\right)\left( \sum_{n=1}^{N_f}\tilde{\sigma}_n^2\right) + \lambda_{6F}''\sum_{m=1}^{N_f}\tilde{\sigma}_m^3\right)\,.
\eeq
We then extremize the total free energy with respect to the $\sigma_m$. (The variations with respect to off-diagonal elements of $\sigma_F$ vanish by the unbroken $SU(N_f)$ flavor symmetry, so it suffices to extremize with respect to its eigenvalues.)

Assuming an $SU(N_f)$-symmetric ansatz for the extremum, i,e. $\sigma_m = \sigma$ for all $m$, we find it to be
\beq
\label{E:Gcritferm}
	\tilde{\sigma}_c = - \text{sgn}(\lambda) \mu_F G\,, \qquad G = \frac{1}{\sqrt{1-2\pi \lambda (N_f^2\lambda_{6F}+N_f \lambda_{6F}'+\lambda_{6F}'')}}\,,
\eeq
which is the same as for a single fermion obtained in~\cite{Aharony:2012ns} upon the replacement $\lambda_{6F} \to N_f^2\lambda_{6F} + N_f \lambda_{6F}'+\lambda_{6F}''$. This expression holds provided that the argument of the square root is positive, and the sign for $\tilde{\sigma}$ is obtained by the requirement that $\mu_F$ is positive. From the equation~\eqref{E:critfermpolemass} for the thermal pole mass, plugging in this value of $\tilde{\sigma}$ we obtain an equation for the pole mass of the critical theory,
\beq
\label{E:eomcritferm}
	(1+G-|\lambda|) \mu_{F,c} =-\frac{1}{\pi i}\left[ \text{Li}_2\left(-e^{-\mu_{F,c}+\pi i |\lambda|}\right)-\text{c.c.}\right]\,.
\eeq
Plugging the extremum $\sigma \to \sigma_c$ into the free energy we obtain after some simplification
\beq
\label{E:Fcritferm}
	\beta F = \frac{NN_f V_2}{2\pi^2i \beta^2|\lambda|} \left( \frac{\mu_{F,c}^2}{3}\text{Li}_2\left( -e^{-\mu_{F,c}+\pi i |\lambda|}\right) + \int_{\mu_{F,c}}^{\infty} dy\,y\, \text{Li}_2\left( -e^{-y+\pi i |\lambda|}\right) - \text{c.c.}\right)\,.
\eeq
This form is ready-made to showcase a duality with the theory with regular scalars under
\beq
	N_f \leftrightarrow N_s\,, \qquad \mu_{F,c} \leftrightarrow \mu_B\,.
\eeq
For this to hold the triple trace couplings of the bosonic and fermionic theories must be appropriately mapped to each other. We discuss this in Section~\ref{S:testRC}.

%***********************************************
\subsection{Critical scalars and regular fermions}
%***********************************************

We obtain the theory with critical scalars and regular fermions by first taking the theory of regular matter, setting the couplings $\lambda_6,\lambda_6'$, etc., to vanish, and then adding a bare bosonic mass $(\sigma_B)^{\alpha}{}_{\beta}$. As above, we diagonalize it, and refer to the eigenvalues as $\sigma_{\alpha}$. There are now $N_s$ thermal pole masses for the bosons, satisfying
\beq
	\sqrt{\mu_{B,\alpha}^2 - \tilde{\sigma}_{\alpha} }= -\frac{1}{2}| \lambda| \mu_{B,\alpha} - \frac{1}{2\pi i}\left( \text{Li}_2\left( e^{-\mu_{B,\alpha}+\pi i |\lambda|}\right)-\text{c.c.}\right)\,.
\eeq
Repeating the analysis for the free energy we find
\beq
	\beta F=  N_F \beta F_f (\mu_F)+\sum_{\alpha=1}^{N_s} \beta F_s(\mu_{B,\alpha},\sigma_{\alpha})\,,
\eeq
where $F_f(\mu_F)$ is the free energy of a single regular fermion with pole mass $\mu_F$, given below, and $F_s$ is the free energy of a single regular scalar with pole mass $\mu_{B}$ and bare mass $\sigma$,
\beq
	\beta F_s(\mu_B,\sigma)=-\frac{NV_2}{2\pi \beta^2 \lambda} \left( \frac{\lambda \mu_B \tilde{\sigma}}{3}+\frac{\mu_{B}^2-\tilde{\sigma}}{3\pi i}\text{Li}_2\left(e^{-\mu_{B}+\pi i \lambda}\right)+\frac{1}{\pi i}\int_{\mu_{B}}^{\infty} dy\,y \,\text{Li}_2\left( e^{-y+\pi i \lambda}\right) - \text{c.c.}\right)\,,
\eeq 
Extremizing the free energy with respect to the $\sigma_{\alpha}$ we find that they are all identical, with
\beq
	\tilde{\sigma}_{\alpha} = \mu_{B,c}^2 \,,
\eeq
and $\mu_{B,c}$ is the thermal pole mass of the critical theory. It and the fermion pole mass are determined by
\begin{align}
\begin{split}
\label{E:masterF1}
	|\lambda| \mu_{B,c} &=\frac{1}{\pi i} \left[ \text{Li}_2\left( e^{-\mu_{B,c} + \pi i| \lambda|}\right)-\text{c.c.}\right]\,,
	\\
	\pm \mu_F & = |\lambda| \mu_F - \frac{1}{\pi i}\left[ \text{Li}_2\left( -e^{-\mu_F+\pi i |\lambda|}\right)-\text{c.c.}\right]\,,
\end{split}
\end{align}
Plugging this back into the free energy we obtain
\beq
\label{E:masterF2}
	\beta F = N_f \beta F_f(\mu_F) + N_s \beta F_{s,c}(\mu_{B,c})\,,
\eeq
where 
\begin{align}
\begin{split}
\label{E:masterF3}
	\beta F_f (\mu_F)&= \frac{NV_2}{2\pi^2i \beta^2|\lambda|}\left( \frac{\mu_F^2}{3}\text{Li}_2\left( -e^{-\mu_F+\pi i |\lambda|}\right) + \int_{\mu_F}^{\infty}dy\,y\,\text{Li}_2\left(-e^{-y+\pi i |\lambda|}\right) - \text{c.c.}\right)\,,
	\\
	\beta F_{s,c}(\mu_{B,c}) &=-\frac{NV_2}{2\pi i \beta^2|\lambda|} \left( \frac{\mu_{B,c}^2}{3}\text{Li}_2\left(e^{-\mu_{B,c}+\pi i |\lambda|}\right) + \int_{\mu_{B,c}}^{\infty} dy\,y\,\text{Li}_2\left( e^{-y+\pi i |\lambda|}\right)-\text{c.c.}\right)\,,
\end{split}
\end{align}
are the free energies of the theories with a single regular fermion or critical boson. 

%***********************************************
\section{Large $N$ zero temperature correlators}
\label{S:correlators}
%***********************************************

$U(N)_k$ Chern-Simons theory coupled to fundamental matter has a simple spectrum of gauge-invariant operators. There are the analogue of ``single trace'' operators, which are matter bilinears with derivatives acting on the matter fields. For example, for Chern-Simons theory coupled to $N_f$ regular fermions and $N_s$ regular scalars, there are scalar bilinears of spin $s=0,1,..$ of the schematic form $J^b_s\sim (\phi^{\dagger} \partial^s\phi)$ and dimension $\Delta = 1+s+O(1/N)$, transforming in the adjoint representation of the $U(N_s)$ that rotates the scalars. There is also a tower of fermion bilinears $J^f_s \sim (\bar{\psi} \partial^s\psi)$ transforming in the adjoint representation of the $U(N_f)$ that rotates the fermions. Those operators are bosonic, but when there are fermions and bosons we also have a tower of fermionic gauge singlets $\sim(\bar{\psi}\partial^s \phi)$. The single-trace operators have dimensions which are $O(1)$ in the `t Hooft limit. The $U(N)_k$ theory also has monopole operators, whose dimensions are parametrically of $O(N)$. There are also ``multi-trace'' operators built from the product of ``single-trace'' operators with each other, and more general products with monopole operators. The $SU(N)_k$ theory does not have monopole operators, but it does have baryon operators whose dimensions are also $O(N)$. 

Building upon previous work~\cite{Aharony:2012nh,GurAri:2012is} we consider the correlation functions of the single-trace operators, focusing on the spin-zero boson and fermion bilinears. We do not consider the mixed scalar/fermion bilinears, though it would be interesting to study these. We decompose the purely scalar and purely fermion bilinears according to their transformation under the $SU(N_s)$ and $SU(N)_f$ global symmetries, i.e. into traceless and traceful parts,
\beq
	 (J_b)^{\alpha}{}_{\beta} = (\phi_{\beta}^{\dagger}\phi^{\alpha}) = (A_b)^{\alpha}{}_{\beta} + \frac{S_b}{N_s} \delta^{\alpha}{}_{\beta}\,, \qquad (J_f)^{m}{}_n = (\bar{\psi}_n \psi^m) = (A_f)^m{}_n + \frac{S_f}{N_f}\delta^m{}_n\,.
\eeq

In writing the matter potential of the theory with regular scalars and regular fermions, or the theory with critical fermions, we did so without performing such a decomposition. However it is useful to do so when computing correlation functions. For the theory with regular scalars and regular fermions we rewrite the approximately marginal triple-trace and double-trace deformations in~\eqref{E:matter} as
\begin{align}
\begin{split}
	V(\phi) &=\tilde{\lambda}_6 S_b^3 +3 \tilde{\lambda}_6' S_b \text{tr}(A_b^2) + \tilde{\lambda}_6'' \text{tr}(A_b^3)\,,
	\\
	U(\phi,\psi) & = \tilde{\lambda}_4 S_b S_f + \tilde{\lambda}_4' \text{tr}(A_b A_f)\,,
\end{split}
\end{align}
with
\begin{align}
\begin{split}
\label{E:reorganized}
	\tilde{\lambda}_6 &= \frac{N_s^2 \lambda_6+N_s\lambda_6'+\lambda_6''}{N_s^2}\,, \qquad \tilde{\lambda}_6' = \frac{\lambda_6'}{3} + \frac{\lambda_6''}{N_s}\,, \qquad \tilde{\lambda}_6'' = \lambda_6''\,,
	\\
	\tilde{\lambda}_4 & = \frac{N_f\lambda_4+\lambda_4'}{N_f}\,, \hspace{0.84in} \tilde{\lambda}_4' = \lambda_4'\,.
\end{split}
\end{align}
Note that the thermal free energy, summarized in Eqs.~\eqref{E:finalFormEOMs} and~\eqref{E:Fregreg}, depends on $\tilde{\lambda}_6$ and $\tilde{\lambda}_4$, but not the other multi-trace couplings. 

Similarly for the theory with critical fermions, described in Subsection~\ref{S:critferm}, we decompose $(\sigma_F)^m{}_n$ as
\beq
	(\sigma_F)^m{}_n = \tilde{A}^m{}_n + \frac{\tilde{S}}{N_f}\delta^m{}_n\,,
\eeq
and so the triple trace couplings as
\beq
	\Delta \mathcal{L} = \frac{N}{3!}\left( \tilde{\lambda}_{6F} \tilde{S}^3 +3 \tilde{\lambda}_{6F}' \tilde{S} \text{tr}(\tilde{A}^2) + \tilde{\lambda}_{6F}''\text{tr}(\tilde{A}^3)\right)\,,
\eeq
with
\beq
\label{E:critfermTripleTrace}
	\tilde{\lambda}_{6F} = \frac{N_f^2 \lambda_{6F}+N_f \lambda_{6F}' + \lambda_{6F}''}{N_f^2}\,, \qquad \tilde{\lambda}_{6F}' = \frac{\lambda_{6F}'}{3} + \frac{\lambda_{6F}''}{N_f} \,, \qquad \tilde{\lambda}_{6F}'' = \lambda_{6F}''\,.
\eeq
Also note that the thermal free energy of this theory, summarized in Eqs.~\eqref{E:eomcritferm} and~\eqref{E:Fcritferm}, only depends on $\tilde{\lambda}_{6F}$ but not $\tilde{\lambda}_{6F}'$ or $\tilde{\lambda}_{6F}''$. 

%***********************************************
\subsection{Two-point functions}
\label{S:twopoint}
%***********************************************

\begin{figure}[t]
\begin{center}
\includegraphics[width=6in]{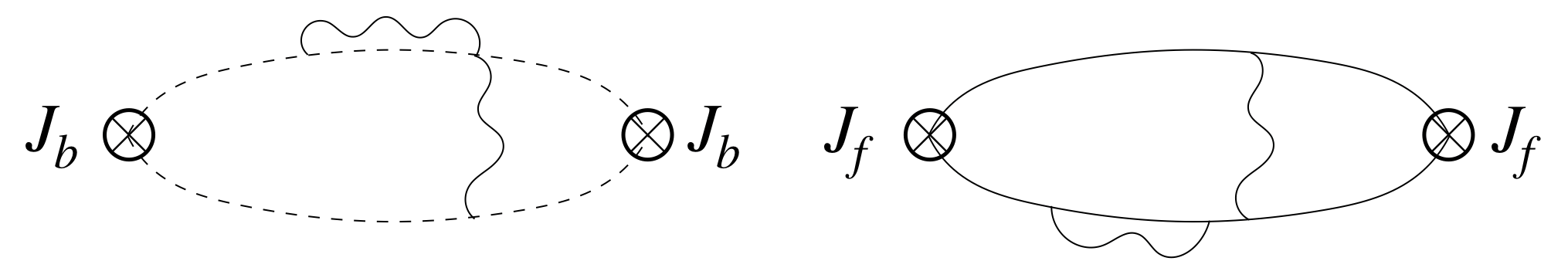}
\end{center}
\caption{\label{F:twopoint1} Some contributions to the large $N$ two-point functions of $J_b$ and $J_f$. Dashed lines indicate the scalar $\phi$, solid the fermion $\psi$, wavy lines the gauge field, crosses insertions of $J_b$ and $J_f$.}
\end{figure} 

Consider $U(N)_k$ Chern-Simons theory coupled to either a single boson of fermion. Denote the scalar bilinear as $J_b = (\phi^{\dagger}\phi)$ and $J_f = (\bar{\psi}\psi)$. The two-point functions of $J_b$ and $J_s$ are~\cite{Aharony:2012nh,GurAri:2012is}
\begin{align}
\begin{split}
	\langle J_b(q)J_b\rangle^{0} & = G_b(q)=\frac{N\tan\left(\frac{\pi |\lambda|}{2}\right)}{4\pi |\lambda|} \frac{1}{|q|}\,,
	\\
	\langle J_f(q)J_f\rangle^{0} & = G_f(q)=-\frac{N\tan\left(\frac{\pi |\lambda|}{2}\right)}{4\pi|\lambda|}|q|\,,
\end{split}
\end{align}
where we work in momentum space and have stripped off the usual factor of $(2\pi)^3 \delta^{(3)}(0)$. The superscript $0$ on the left-hand-side indicates that these correlators are evaluated in the theory with only bosons or fermions. These correlators are the result of summing up diagrams of the sort pictured in Fig.~\ref{F:twopoint1}, where gauge bosons are exchanged between exact large $N$ boson or fermion propagators. These diagrams are insensitive to the number of flavors, and so we immediately obtain the two-point functions of $(J_b)^{\alpha}{}_{\beta}$ and $(J_m)^m{}_n$ in $U(N)_k$ Chern-Simons theory coupled to $N_s$ scalars or $N_f$ fermions:
\begin{align}
\begin{split}
	\langle (J_b)^{\alpha_1}{}_{\alpha_2}(q) (J_b)^{\alpha_3}{}_{\alpha_4}\rangle^0 & =  G_b(q)\delta^{\alpha_1}{}_{\alpha_4}\delta^{\alpha_3}{}_{\alpha_2}\,,
	\\
	\langle (J_f)^{m_1}{}_{m_2}(q)(J_f)^{m_3}{}_{m_4}\rangle^0 & = G_f(q)\delta^{m_1}{}_{m_4}\delta^{m_2}{}_{m_3}\,.
\end{split}
\end{align}
Decomposing into the adjoint and scalar parts, we find
\begin{align}
\begin{split}
\label{E:twopointstarting}
	\langle (A_b)^{\alpha_1}{}_{\alpha_2}(q)(A_b)^{\alpha_3}{}_{\alpha_4}\rangle^0 & = G_b(q)(I_b)^{\alpha_1\alpha_3}_{\alpha_2\alpha_4}\,, \qquad\hspace{.11in} \langle S_b(q)S_b\rangle^0 = N_s G_b(q)\,, 
	\\
	\langle (A_f)^{m_1}{}_{m_2}(q)(A_f)^{m_3}{}_{m_4}\rangle^0 &= G_f(q) (I_f)^{m_1m_3}_{m_2m_4}\,,\qquad \langle S_f(q)S_f\rangle^0 = N_f G_f(q)\,,
\end{split}
\end{align}
where
\beq
	(I_b)^{\alpha_1\alpha_3}_{\alpha_2\alpha_4} = \delta^{\alpha_1}{}_{\alpha_4}\delta^{\alpha_3}{}_{\alpha_2}-\frac{1}{N_s}\delta^{\alpha_1}{}_{\alpha_2}\delta^{\alpha_3}{}_{\alpha_4}\,,
\eeq
is the unique flavor structure allowed in the two-point function of fields in the adjoint representation. These two-point functions are the foundation we build upon. 

Now consider $U(N)_k$ Chern-Simons theory coupled to $N_f$ fermions and $N_s$ scalars with no double-trace interactions, $\tilde{\lambda}_4 = \tilde{\lambda}_4' = 0$. In this theory the fermions and scalars are not coupled to leading order in large $N$, and the large $N$ two-point function of $J_b$ and $J_f$ are as above. Turning on $\tilde{\lambda}_4$ and $\tilde{\lambda}_4'$ then amounts to deforming by double-trace deformations, which lead to a simple modification of two- and three-point functions at large $N$.

We could also study the two-point functions of the spin$-s$ tower of scalar and fermion bilinears. However these are unaltered by the matter couplings discussed below, and so we do not consider them further.

%***********************************************
\subsubsection{Regular scalars and fermions}
\label{S:2ptregreg}
%***********************************************

Now consider the theory with $N_s$ regular scalars and $N_f$ regular fermions, which we view as a double trace deformation by $\tilde{\lambda}_4$ and $\tilde{\lambda}_4'$.  \newline

\begin{figure}[t]
\begin{center}
\includegraphics[width=3.7in]{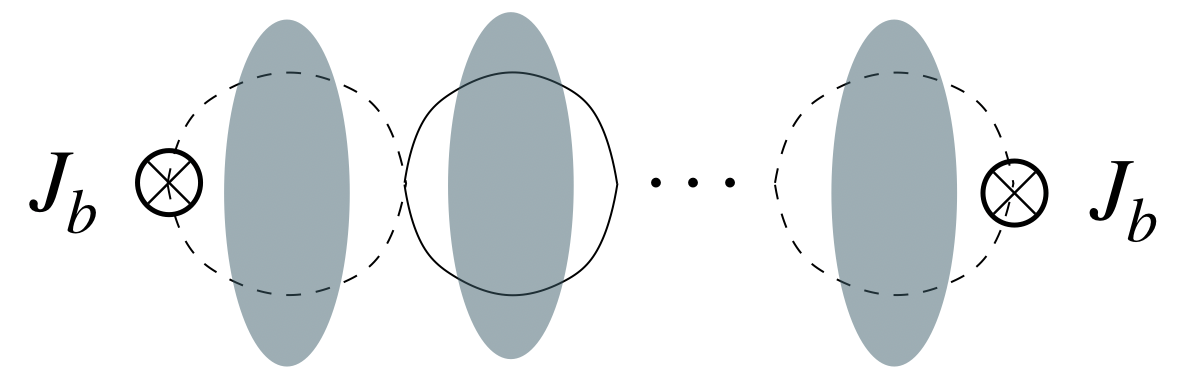}
\end{center}
\caption{\label{F:twopoint2} The two-point function of $J_b$ in the double-trace deformed theory is a sum of bubbles involving the two-point function of $J_b$ in the undeformed theory and that of $J_f$, denoted by blobs.}
\end{figure}

\noindent \underline{$N_s \neq N_F$}: In this case there is a unique coupling which contributes to the large $N$ two-point functions, $\sim \tilde{\lambda}_4 S_b S_f$. This in turn modifies the two-point functions of the flavor scalars $S_b$ and $S_f$, which are now given by a sum of bubble diagrams as pictured in Fig.~\ref{F:twopoint2}. The two-point functions of adjoints are unmodified, while the sum of bubbles gives the geometric series
\begin{subequations}
\label{E:twoptregreg1}
\begin{align}
\begin{split}
	\langle S_b(q) S_b\rangle &=\langle S_b(q)S_b\rangle^0  \sum_{n=0}^{\infty} \left[ \left( -\frac{\tilde{\lambda}_4}{N}\right)^2 \langle S_f(q)S_f\rangle^0 \langle S_b(q)S_b\rangle^0\right]^n
	\\
	& = \frac{N_s G_b(q)}{1+x_4^2N_s N_f \tan^2\left( \frac{\pi \lambda}{2}\right)}\,, \qquad x_4 = \frac{\tilde{\lambda}_4}{4\pi \lambda} \,.
\end{split}
\end{align}
Similarly, defining the bubble function
\beq
	B_S = \frac{1}{1+x_4^2N_s N_f \tan^2\left(\frac{\pi \lambda}{2}\right)}\,,
\eeq
we have
\beq
	\langle S_f(q)S_f\rangle = N_fB_S G_f(q)\,.
\eeq
We also find that the two-point function of $S_b$ with $S_f$ is nonzero, albeit a pure contact term,
\beq
	\langle S_b(q)S_f\rangle = -x_4 N_s N_f B_S\tan^2\left( \frac{\pi \lambda}{2}\right)\,.
\eeq
\end{subequations}

\noindent \underline{$N_s=N_f$}: There are now two couplings which contribute to the large $N$ two-point functions. The two-point functions of the scalars $S_b$ and $S_f$ are given by the expressions~\eqref{E:twoptregreg1}. The two-point functions of the adjoints are modified as
\begin{align}
\begin{split}
	\langle (A_b)^{m_1}{}_{m_2}(q)(A_b)^{m_3}{}_{m_4}\rangle & = B_AG_b(q)(I_b)^{m_1m_3}_{m_2m_4}\,,
	\\
	\langle (A_f)^{m_1}{}_{m_2}(q)(A_f)^{m_3}{}_{m_4}\rangle &= B_AG_f(q)(I_b)^{m_1m_3}_{m_2m_4}\,, 
	\\
	\langle (A_b)^{m_1}{}_{m_2}(q)(A_f)^{m_3}{}_{m_4}\rangle & = -x_4' \tan^2\left(\frac{\pi \lambda}{2}\right)B_A(I_b)^{m_1m_3}_{m_2m_4}\,,
\end{split}
\end{align}
with
\beq
	x_4' = \frac{\tilde{\lambda}_4'}{4\pi\lambda}\,,
\eeq
and where we have defined the bubble function
\beq
	B_A = \frac{1}{1+x_4'^2 \tan^2\left(\frac{\pi \lambda}{2}\right)}\,.
\eeq

%***********************************************
\subsubsection{Regular scalars or critical fermions}
\label{S:2ptcritferm}
%***********************************************

The theory with $N_s$ regular scalars is already subsumed in our analysis above. The two-point functions of $A_b$ and $S_b$ are given by~\eqref{E:twopointstarting}. 

To get to the theory of $N_f$ critical fermions, we begin with the theory of $N_f$ regular massless fermions. The critical theory is obtained by Legendre transforming with respect to the fermion bilinear $J_f$. The two-point function of the Legendre dual $\sigma_F$ is
\begin{align}
\begin{split}
	\langle (\sigma_F)^{m_1}{}_{m_2}(q) (\sigma_F)^{m_3}{}_{m_4}\rangle &= \left( -G_f(q)\right)^{-1}\delta^{m_1}{}_{m_4}\delta^{m_3}{}_{m_2}  = G_{f,c}(q)\delta^{m_1}{}_{m_4}\delta^{m_3}{}_{m_2}\,,
	\\
	G_{f,c}(q)&= \frac{4\pi |\lambda|\cot\left(\frac{\pi |\lambda|}{2}\right)}{N}\frac{1}{|q|}\,.
\end{split}
\end{align}
Decomposing $\sigma_F$ into traceless and traceful parts $\tilde{A}$ and $\tilde{S}$, we have
\beq
	\langle \tilde{A}{m_1}{}_{m_2}(q)\tilde{A}^{m_3}{}_{m_4}\rangle = G_{f,c}(q) (I_f)^{m_1m_3}_{m_2m_4}\,, \qquad \langle \tilde{S}(q)\tilde{S}\rangle = N_f G_{f,c}(q)\,.
\eeq

%***********************************************
\subsubsection{Critical scalars and regular fermions}
%***********************************************

This theory is obtained by starting with the theory of $N_s$ regular scalars and $N_f$ regular fermions, setting the sextic and quartic deformations $\tilde{\lambda}_6,\tilde{\lambda}_6'$, etc., to vanish. We then Legendre transform with respect to the scalar bilinear $J_b$. Denoting the dual field as $\tilde{J}_b$, we have
\begin{align}
\begin{split}
	\langle (\tilde{J}_b)^{\alpha_1}{}_{\alpha_2}(q)(\tilde{J}_b)^{\alpha_3}{}_{\alpha_4}\rangle  & =  \left( - G_b(q)\right)^{-1} \delta^{\alpha_1}{}_{\alpha_4}\delta^{\alpha_3}{}_{\alpha_2} = G_{b,c}(q) \delta^{\alpha_1}{}_{\alpha_4}\delta^{\alpha_3}{}_{\alpha_2}\,,
	\\
	G_{b,c}(q) &= - \frac{4\pi |\lambda|\cot\left( \frac{\pi |\lambda|}{2}\right)}{N}|q|\,.
\end{split}
\end{align}
Decomposing $\tilde{J}_b$ into traceless and traceful parts $\tilde{A}_b$ and $\tilde{S}_b$, we have
\beq
	\langle (\tilde{A}_b)^{\alpha_1}{}_{\alpha_2}(q)(\tilde{A}_b)^{\alpha_3}{}_{\alpha_4}\rangle = G_{b,c}(q) (I_b)^{\alpha_1\alpha_3}_{\alpha_2\alpha_4}\,, \qquad \langle \tilde{S}_b(q)\tilde{S}_b\rangle \rangle = N_sG_{b,c}(q)\,.
\eeq
Meanwhile the two-point functions of fermions are those of the pure fermion theory~\eqref{E:twopointstarting}. 

%***********************************************
\subsection{Three-point functions}
\label{S:threepoint}
%***********************************************

\begin{figure}[t]
\begin{center}
\includegraphics[width=3in]{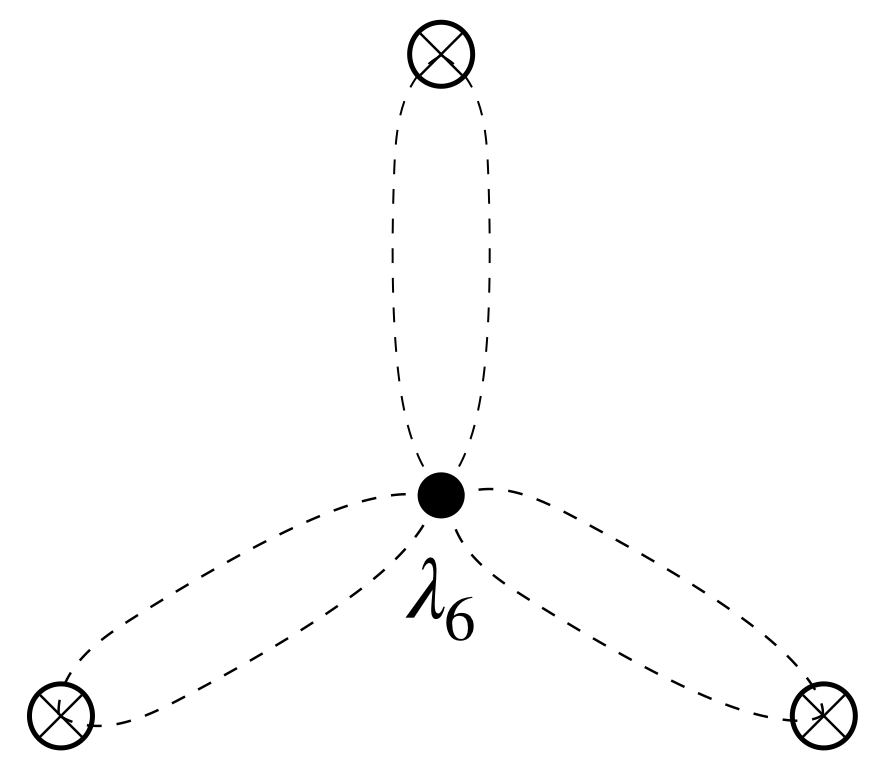}
\end{center}
\caption{\label{F:lambda6} The contribution of the sextic coupling $\lambda_6$ to the three-point function of $J_b$. There are implicit blobs connecting adjacent scalar propagators, which sum up the gauge interactions at large $N$.}
\end{figure}

Consider again the three-point functions of $J_b$ or $J_f$ in $U(N)_k$ Chern-Simons theory coupled to either a single regular boson or regular fermion. The result of~\cite{Aharony:2012nh,GurAri:2012is} is
\begin{align}
\begin{split}
\label{E:starting3pt}
	\langle J_b(q)J_b(q')J_b\rangle^0 & = \frac{N}{2\pi \lambda}\left( \frac{\tan\left(\frac{\pi \lambda}{2}\right)}{\cos^2\left(\frac{\pi \lambda}{2}\right)} - \frac{\tan^3\left(\frac{\pi \lambda}{3}\right)}{4}\left( 1+ \frac{\lambda_6}{8\pi^2 \lambda^2}\right)\right)\frac{1}{|q||q'||q+q'|}\,,
	\\
	\langle J_f(q)J_f(q')J_f\rangle^0 & = \frac{N}{2\pi \lambda}\tan^2\left(\frac{\pi \lambda}{2}\right)\,.
\end{split}
\end{align}
Note that the three-point function of $J_b$ is of the correct form for a dimension $1$ primary, while the three-point function of $J_f$ is a pure contact term. The dependence of the three-point function of $J_b$ on the sextic coupling $\lambda_6$ only arises through the diagram in Fig.~\ref{F:lambda6}, which factorizes as $-\frac{\lambda_6}{N^2} \langle J_b(q)J_b\rangle^0 \langle J_b(q')J_b\rangle^0\langle J_b(-q-q')J_b\rangle^0$. 

The generalization to $U(N)_k$ Chern-Simons theory coupled to $N_s$ scalars is
\beq
	\langle (J_b)^{\alpha_1}{}_{\alpha_2}(q) (J_b)^{\alpha_3}{}_{\alpha_4}(q')(J_b)^{\alpha_5}{}_{\alpha_6}\rangle^0 = T_b(q,q') P^{\alpha_1\alpha_3\alpha_5}_{\alpha_2\alpha_4\alpha_6} + T_b'(q,q')(P')^{\alpha_1\alpha_3\alpha_5}_{\alpha_2\alpha_4\alpha_6} + T_b''(P'')^{\alpha_1\alpha_3\alpha_5}_{\alpha_2\alpha_4\alpha_6}\,,
\eeq
with
\begin{align}
\begin{split}
	T_b& = \frac{N}{2\pi \lambda}\left( \frac{\tan\left(\frac{\pi \lambda}{2}\right)}{\cos^2\left(\frac{\pi \lambda}{2}\right)} - \frac{\tan^3\left(\frac{\pi \lambda}{2}\right)}{4}\left(1+\frac{\lambda_6''}{8\pi^2 \lambda^2}\right)\right)\frac{1}{|q||q'||q+q'|}\,,
	\\
	T_b' & = -\frac{N\lambda_6'\tan^3\left(\frac{\pi \lambda}{2}\right)}{64\pi^3\lambda^3} \frac{1}{|q||q'||q+q'|}\,,
	\\
	T_b'' &= -\frac{N\lambda_6\tan^3\left(\frac{\pi \lambda}{2}\right)}{64\pi^3\lambda^3} \frac{1}{|q||q'||q+q'|}\,,
\end{split}
\end{align}
and the tensor structures are
\begin{align}
\begin{split}
	P^{\alpha_1\alpha_3\alpha_5}_{\alpha_2\alpha_4\alpha_6} &= \frac{1}{3!}\left( \delta^{\alpha_1}{}_{\alpha_4}\delta^{\alpha_3}{}_{\alpha_6}\delta^{\alpha_5}{}_{\alpha_2}+(\text{permutations})\right)\,,
	\\
	(P')^{\alpha_1\alpha_3\alpha_5}_{\alpha_2\alpha_4\alpha_6} & = \frac{1}{3}\left( \delta^{\alpha_1}{}_{\alpha_2}\delta^{\alpha_3}{}_{\alpha_6} \delta^{\alpha_5}{}_{\alpha_4} + \delta^{\alpha_3}{}_{\alpha_4}\delta^{\alpha_1}{}_{\alpha_6}\delta^{\alpha_5}{}_{\alpha_2} + \delta^{\alpha_5}{}_{\alpha_6}\delta^{\alpha_1}\delta_{\alpha_4}\delta^{\alpha_3}{}_{\alpha_2}\right)\,,
	\\
	(P'')^{\alpha_1\alpha_3\alpha_5}_{\alpha_2\alpha_4\alpha_6} &= \delta^{\alpha_1}{}_{\alpha_2}\delta^{\alpha_3}{}_{\alpha_4}\delta^{\alpha_5}{}_{\alpha_6}\,.
\end{split}
\end{align}
This data can be reorganized into the three-point functions of the traceless and traceful parts of $J_b$. The three-point function of adjoints is essentially that for the theory of a single scalar,
\beq
\label{E:3ptscalar1}
	\langle (A_b)^{\alpha_1}{}_{\alpha_2}(q)(A_b)^{\alpha_3}{}_{\alpha_4}(q')(A_b)^{\alpha_5}{}_{\alpha_6}\rangle^0 = T_b(q,q') (H_b)^{\alpha_1\alpha_3\alpha_5}_{\alpha_2\alpha_4\alpha_6} \,,
\eeq
up to a group theory factor
\beq
	(H_b)^{\alpha_1\alpha_3\alpha_5}_{\alpha_2\alpha_4\alpha_6}=\frac{1}{3!}\left( \delta^{\alpha_1}{}_{\alpha_4}\delta^{\alpha_3}{}_{\alpha_6}\delta^{\alpha_5}{}_{\alpha_2}+(\text{permutations}) - (\text{traces})\right)\,,
\eeq
appropriate for the three-point function of adjoint representations.
We also have
\begin{align}
\label{E:3ptscalar2}
	\langle S(q) (A_b)^{\alpha_1}{}_{\alpha_2}(q)(A_b)^{\alpha_3}{}_{\alpha_4}\rangle^0 &=\left( T_b(q,q')+\frac{N_s}{3}T_b'(q,q')\right)(I_b)^{\alpha_1\alpha_3}_{\alpha_2\alpha_4} 
	\\
	\nonumber
	&= \frac{N}{2\pi \lambda}\left(\frac{\tan\left(\frac{\pi\lambda}{2}\right)}{\cos^2\left(\frac{\pi \lambda}{2}\right)} - \frac{\tan^3\left( \frac{\pi \lambda}{2}\right)}{4}\left( 1 + \frac{ N_s \tilde{\lambda}_6'}{8\pi^2\lambda^2}\right)\right)\frac{1}{|q||q'||q+q'|} (I_b)^{\alpha_1\alpha_3}_{\alpha_2\alpha_4}\,,
\end{align}
where $\tilde{\lambda}_4'$ is the redefined sextic coupling $\sim \tilde{\lambda}_6'S_b \text{tr}(A_b^2)$ given in~\eqref{E:reorganized}. Finally the three-point function of traces is
\begin{align}
\begin{split}
\label{E:3ptscalar3}
	\langle S(q)S(q')S\rangle^0 & = N_s T_b(q,q')+N_s^2 T_b'(q,q') + N_s^3 T_b''(q,q')
	\\
	& = \frac{N N_s}{2\pi \lambda}\left( \frac{\tan\left(\frac{\pi \lambda}{2}\right)}{\cos^2\left(\frac{\pi \lambda}{2}\right)} - \frac{\tan^3\left(\frac{\pi \lambda}{2}\right)}{4}\left( 1 +\frac{N_s^2\tilde{\lambda}_6}{8\pi^2\lambda^2}\right)\right)\frac{1}{|q||q'||q+q'|} \,,
\end{split}
\end{align}
where $\tilde{\lambda_6}$ is the sextic coupling $\sim \tilde{\lambda}_6S_b^3$ in~\eqref{E:reorganized}. The three-point function of a single $A_b$ with two scalars $S_b$ is forbidden by symmetry.

In each case we see that the dependence of the three-point function on the couplings is the same as that for a single scalar~\eqref{E:starting3pt} upon replacing the triple-trace coupling $\lambda_6$ with either the triple-trace coupling of three adjoints, $\tilde{\lambda}_6''=\lambda_6''$, the coupling of a scalar with two adjoints, $\tilde{\lambda}_4'$, or the coupling of three scalars, $\tilde{\lambda}_6$. With this in mind, we define the normalization factor
\beq
	\mathcal{N}_b(z) = \frac{N}{2\pi \lambda}\left( \tan\left( \frac{\pi\lambda}{2}\right) -z\tan^3\left(\frac{\pi \lambda}{2}\right)\right) \,,
\eeq
along with
\beq
\label{E:y}
	z_6 = \frac{\hat{\lambda}^2}{4\lambda^2}-1 = \frac{1}{4}\left(-3+\frac{N_s^2 \tilde{\lambda}_6}{8\pi^2\lambda^2}\right)\,, \,\,\, z_6' = \frac{1}{4}\left(-3+\frac{N_s\tilde{\lambda}_6'}{8\pi^2\lambda^2}\right)\,, \,\,\,z_6'' = \frac{1}{4}\left(-3+\frac{\tilde{\lambda}_6''}{8\pi^2\lambda^2}\right)\,.
\eeq
The combination $\frac{\hat{\lambda}^2}{4\lambda^2}$ is the same one that appears in the thermal free energy~\eqref{E:finalFormEOMs} of this model. In this convention, the three-point functions are
\begin{align}
\begin{split}
\label{E:3ptscalar2}
	\langle(A_b)^{\alpha_1}{}_{\alpha_2}(q) (A_b)^{\alpha_3}{}_{\alpha_4}(q')(A_b)^{\alpha_5}{}_{\alpha_6}\rangle & = \mathcal{N}_b(z_6'') \frac{(H_b)^{\alpha_1\alpha_3\alpha_5}_{\alpha_2\alpha_4\alpha_6}}{|q||q'||q+q'|}\,,
	\\
	\langle S_b(q)(A_b)^{\alpha_1}{}_{\alpha_2}(q')(A_b)^{\alpha_3}{}_{\alpha_4}\rangle & = \mathcal{N}_b(z_6') \frac{(I_b)^{\alpha_1\alpha_3}_{\alpha_2\alpha_4}}{|q||q'||q+q'|}\,,
	\\
	\langle S_b(q)S_b(q')S_b\rangle & = \frac{N_s \mathcal{N}_b(z_6)}{|q||q'||q+q'|}\,.
\end{split}
\end{align}
Similarly, for $U(N)_k$ Chern-Simons theory coupled to $N_f$ fermions we have
\beq
	\langle (J_f)^{m_1}{}_{m_2}(q)(J_f)^{m_3}{}_{m_4}(q')(J_f)^{m_5}{}_{m_6})\rangle^0 = T_f(q,q') P^{m_1m_3m_5}_{m_2m_4m_6}\,, 
\eeq
with
\beq
	T_f(q,q') = \frac{N}{2\pi\lambda}\tan^2\left(\frac{\pi \lambda}{2}\right)\,.
\eeq
Decomposing into the three-point function of adjoints and scalars we have
\begin{align}
\begin{split}
\label{E:3ptfermion1}
	\langle (A_f)^{m_1}{}_{m_2}(q)(A_f)^{m_3}{}_{m_4}(q')(A_f)^{m_5}{}_{m_6}\rangle^0 & = T_f(q,q') (H_f)^{m_1m_3m_5}_{m_2m_4m_6}\,,
	\\
	\langle S_f(q)(A_f)^{m_1}{}_{m_2}(q')(A_f)^{m_3}{}_{m_4}\rangle^0 & = T_f(q,q') (I_f)^{m_1m_3}_{m_2m_4}\,,
	\\
	\langle S_f(q)S_f(q')S_f\rangle^0 &= N_f T_f(q,q')\,.
\end{split}
\end{align}
We also define the (at this stage, redundant) fermionic normalization
\beq
	\mathcal{N}_f = \frac{N}{2\pi\lambda}\tan^2\left(\frac{\pi \lambda}{2}\right)\,.
\eeq

As in our discussion of two-point functions, these results for the three-point functions for the theory of just fermions or just scalars continue to hold for the theory with fermions and scalars, provided that the double-trace couplings $\sim S_b S_f$ and, for $N_s = N_f$, $\text{tr}(A_b A_f)$, are set to vanish. 

%***********************************************
\subsubsection{Regular scalars and fermions}
\label{S:3ptregreg}
%***********************************************

We return to the theory with $N_s$ regular scalars and $N_f$ regular fermions. We turn on the double trace deformations $\tilde{\lambda}_4$ and $\tilde{\lambda}_4'$, and obtain the three-point functions of the deformed theory by summing up bubbles as we did for two-point functions. \newline

\noindent \underline{$N_s\neq N_f$}: As we saw in our analysis of two-point functions, the two-point function of adjoints $A_b$ and $A_f$ are unmodified, but the two-point functions of the scalar bilinear $S_b = (\phi^{\dagger}_{\alpha}\phi^{\alpha})$ and $S_f = (\bar{\psi}_m\psi^m)$ are, by the coupling $\lambda_4 S_b S_f$. In particular $S_b$ can bubble into $S_f$. 

Consequently the three-point functions of adjoints $A_b$ and $A_f$ are unmodified, taking on the values given in Eqs.~\eqref{E:3ptscalar1},~\eqref{E:3ptscalar2},~\eqref{E:3ptscalar3}, and~\eqref{E:3ptfermion1}. The mixed three-point functions of $A_b$ with $A_f$ vanish.

\begin{figure}[t]
\begin{center}
\includegraphics[width=3.5in]{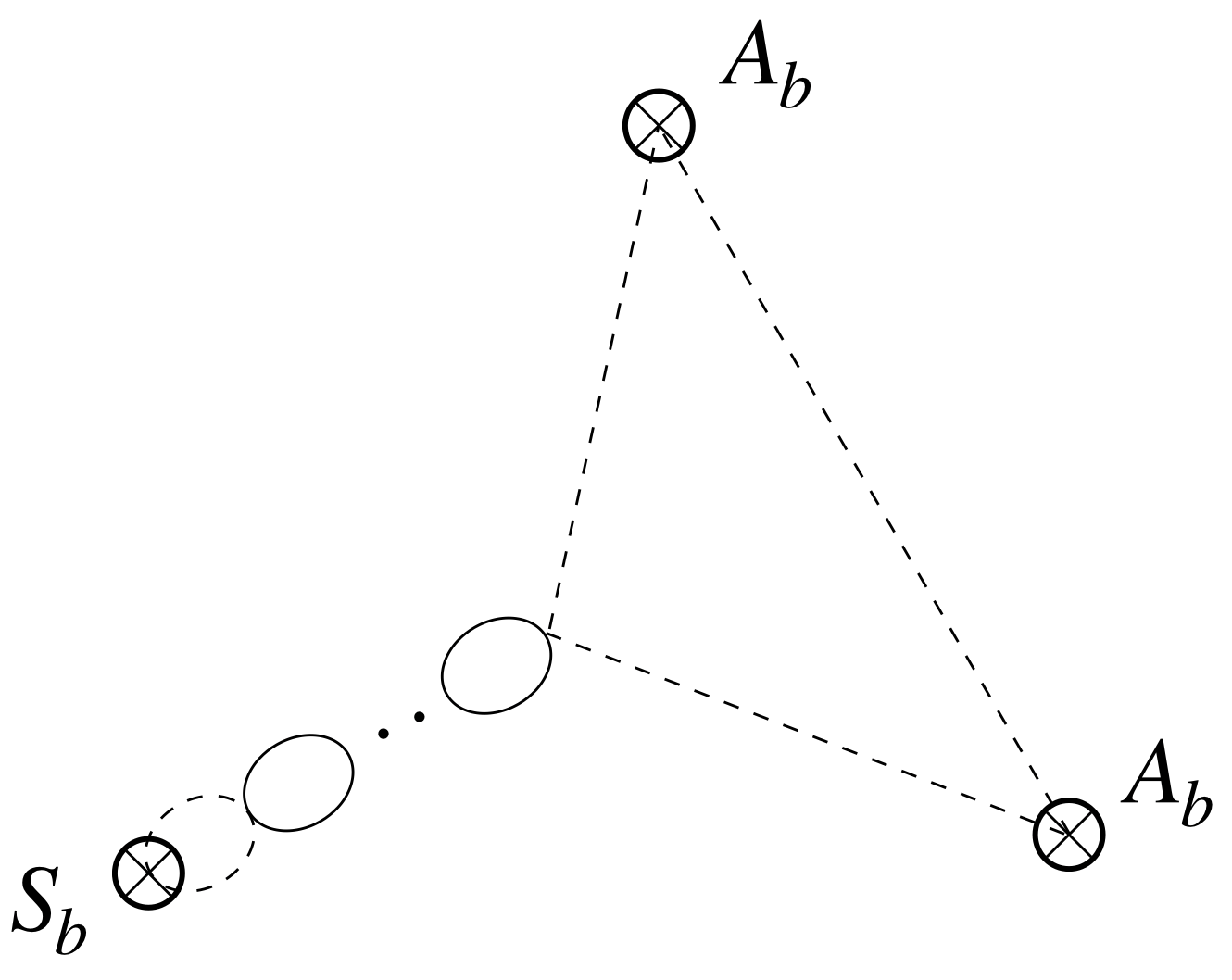}
\end{center}
\caption{\label{F:3pt1} The three-point function of $S_b$ with two $A_b$'s is that of the undeformed theory, with a sum over bubbles attached to the $S_b$ vertex of the undeformed three-point function.}
\end{figure}

The three-point function of $S_b$ with two $A_b$'s is now modified. The large $N$ correlator is given by gluing a sum of bubble diagrams to the $S_b$ vertex of the three-point function $\langle S_b A_b A_b\rangle^0$ in~\eqref{E:3ptscalar2}. See Fig.~\ref{F:3pt1}. The result is
\begin{align}
\nonumber
	\langle S_b(q)(A_b)^{\alpha_1}{}_{\alpha_2}(q')(A_b)^{\alpha_3}{}_{\alpha_4}\rangle& = \sum_{n=0}^{\infty}\left[\left(- \frac{\tilde{\lambda}_4}{N}\right)^2 \langle S_b(q)S_b\rangle^0\langle S_f(q)S_f\rangle^0\right]^n \langle S_b(q)(A_b)^{\alpha_1}{}_{\alpha_2}(q')(A_b)^{\alpha_3}{}_{\alpha_4}\rangle^0
	\\
	& = B_S\mathcal{N}_b(z_6')(I_b)^{\alpha_1\alpha_3}_{\alpha_2\alpha_4}\,.
\end{align}
(Here $x_4 = \frac{\tilde{\lambda}_4}{4\pi \lambda}$.) There is a similar modification to the three-point function of $S_f$ with two $A_f$'s, although since this is a pure contact term we neglect it here. The more interesting effect is that we can have a three-point function of $S_f$ with two $A_b$'s,
\begin{align}
\begin{split}
	\langle S_f(q)(A_b)^{\alpha_1}{}_{\alpha_2}(q')(A_b)^{\alpha_3}{}_{\alpha_4}\rangle & =   x_4N_fB_S\tan\left(\frac{\pi\lambda}{2}\right)  \mathcal{N}_b(z_6')\frac{(I_b)^{\alpha_1\alpha_3}_{\alpha_2\alpha_4}}{|q'||q+q'|}\,.
\end{split}
\end{align}
This is the appropriate forms for the three-point function of two dimension $1$ operators with a dimension $2$ operators. The three-point functions $\langle S_b A_f A_f\rangle$ and $\langle S_b S_f S_f\rangle$ can also be computed, but these are pure contact terms.

There are also the three-point functions of $S_b$ and $S_f$. The three-point function of $S_b$ is given by a sum of two terms, one involving the three-point function $\langle S_bS_bS_b\rangle^0$ dressed by bubble on each vertex, and another involving the three-point function $\langle S_f S_f S_f\rangle^0$, also dressed by bubbles on each vertex. In an equation,
\begin{align}
\begin{split}
\label{E:3ptscalarsum}
	\langle S_b(q)S_b(q')S_b\rangle & = \left(\sum_{n=0}^{\infty}\left[ \left(\frac{\tilde{\lambda}_4}{N}\right)^2 \langle S_b(q)S_b\rangle^0\langle S_f(q)S_f\rangle^0\right]^n\right)^3 \langle S_b(q)S_b(q')S_b\rangle^0 
	\\
	&\qquad + \mathcal{G}(q)\mathcal{G}(q')\mathcal{G}(-q-q')\langle S_f(q)S_f(q')S_f\rangle^0\,,
\end{split}
\end{align}
with
\begin{align}
\begin{split}
	\mathcal{G}(q) &= -\frac{\tilde{\lambda}_4}{N}\langle S_b(q)S_b\rangle^0\sum_{n=0}^{\infty} \left[ \left(\frac{\tilde{\lambda}_4}{N}\right)^2\langle S_b(q)S_b\rangle^0\langle S_f(q)S_f\rangle^0\right]^n
	\\
	&= -\frac{x_4 N_s \tan\left(\frac{\pi\lambda}{2}\right)}{1+x_4^2N_fN_s \tan^2\left(\frac{\pi\lambda}{2}\right)}\frac{1}{|q|} = -x_4 \langle S_b(q)S_b\rangle\,.
\end{split}
\end{align}
This evaluates to
\beq
	\langle S_b(q)S_b(q')S_b\rangle  = B_S^3\left( N_s\mathcal{N}_s(z_6)-x_4^3 N_s^3 \tan^3\left( \frac{\pi \lambda}{2}\right)N_f\mathcal{N}_f\right)\frac{1}{|q||q'||q+q'|}\,.
\eeq
The mixed three-point function $\langle S_b S_b S_f\rangle$ can be obtained in the same way.\newline

\noindent \underline{$N_s=N_f$}: Since now the adjoint bilinears $A_b$ and $A_f$ can bubble into each other, all three-point functions allowed by symmetry are generated. They arise in essentially the same way as the modified three-point function of $S_b$ above in~\eqref{E:3ptscalarsum}. We find that the three-point functions of $A_b$ and $A_f$ are
\begin{align}
\begin{split}
	\langle (A_b)^{m_1}{}_{m_2}(q)(A_b)^{m_3}{}_{m_4}(q')(A_b)^{m_5}{}_{m_6}\rangle & = B_A^3\left(\mathcal{N}_b(z_6'')-X'^3 \mathcal{N}_f\right)\frac{(H_b)^{m_1m_3m_5}_{m_2m_4m_6}}{|q||q'||q+q'|}\,,
	\\
	\langle (A_b)^{m_1}{}_{m_2}(q) (A_b)^{m_3}{}_{m_4}(q')(A_f)^{m_5}{}_{m_6}\rangle &= B_A^3 \left(X'\mathcal{N}_b(y_6'') +X'^2\mathcal{N}_f\right) \frac{(H_b)^{m_1m_3m_5}_{m_2m_4m_6}}{|q||q'|}\,.
\end{split}
\end{align}
The three-point functions $\langle A_b A_f A_f\rangle$ and $\langle A_f A_fA_f\rangle$ are contact terms. To simplify the expression, we have defined the factor 
\beq
	X' = x_4' \tan\left(\frac{\pi \lambda}{2}\right)\,.
\eeq
The simple way to understand this is the following. Each three-point function is the sum of two terms, one involving the three-point function of adjoints in the pure bosonic theory, $\mathcal{N}_b(\tilde{y}_6'')$, and one involving the three-point function of adjoints in the pure fermion theory, $\mathcal{N}_f$. The group theory factor and momentum dependence are fixed by symmetry. The bubble factors arise for each adjoint insertion. When $A_b$ bubbles into $A_f$ it introduces a factor of $-X'$, and when $A_f$ bubbles into $A_b$ it introduces a factor of $X'$.

The mixed three-point functions of one scalar and two adjoints arise in a similar way,
\begin{align}
\begin{split}
	\langle S_b(q)(A_b)^{m_1}{}_{m_2}(q')(A_b)^{m_3}{}_{m_4}\rangle & = B_SB_A^2 \left(\mathcal{N}_b(z_6') - XX'^2\mathcal{N}_f\right)\frac{(I_b)^{m_1m_3}_{m_2m_4}}{|q||q'||q+q'|}\,,
	\\
	\langle S_f(q)(A_b)^{m_1}{}_{m_2}(q')(A_b)^{m_3}{}_{m_4}\rangle & = B_SB_A^2 \left( X\mathcal{N}_b(z_6') + X'^2\mathcal{N}_f\right)\frac{(I_b)^{m_1m_3}_{m_2m_4}}{|q'||q+q'|}\,,
	\\
	\langle S_b(q)(A_b)^{m_1}{}_{m_2}(q')(A_f)^{m_3}{}_{m_4}\rangle & = B_S B_A^2 \left( X' \mathcal{N}_b(z_6') + XX' \mathcal{N}_f\right)\frac{(I_b)^{m_1m_3}_{m_2m_4}}{|q||q'|}\,,
\end{split}
\end{align}
with $\langle S_b A_f A_f\rangle$, $\langle S_f A_f A_b\rangle$, and $\langle S_f A_f A_f\rangle$ contact terms. In analogy with $X'$ we have defined the factor 
\beq
	X = x_4 N_f \tan\left(\frac{\pi \lambda}{2}\right)\,,
\eeq
which arises whenever $S_f$ bubbles into $S_b$; and minus this factor arises when $S_b$ bubbles into $S_f$. Finally the three-point functions of $SU(N_f)$ scalars are
\begin{align}
\begin{split}
\label{E:regreg3ptSSS}
	\langle S_b(q)S_b(q')S_b\rangle & = N_fB_S^3 \left( \mathcal{N}_b(z_6) - X^3\mathcal{N}_f\right)\frac{1}{|q||q'||q+q'|}\,,
	\\
	\langle S_b(q)S_b(q')S_f\rangle & = N_f B_S^3 \left(X\mathcal{N}_b(z_6) + X^2\mathcal{N}_f\right) \frac{1}{|q||q'|}\,,
\end{split}
\end{align}
with $\langle S_b S_f S_f\rangle$ and $\langle S_f S_f S_f\rangle$ contact terms.

%***********************************************
\subsubsection{Regular scalars or critical fermions}
\label{S:3ptcritferm}
%***********************************************

We already dealt with the theory of $N_s$ regular scalars above. For the theory of $N_f$ critical fermions, we begin with the three-point functions of the theory of $N_f$ regular fermions. Our analysis continues that of Subsection~\ref{S:2ptcritferm}. 

We begin by rescaling the critical fermion ``bilinear $\sigma_F$ as
\beq
	\sigma_F \to \frac{N}{4\pi \lambda}\sigma_F\,,
\eeq
so that the two-point functions of its scalar and adjoint parts $\tilde{S}$ and $\tilde{A}$ are
\beq
\label{E:critferm2pt}
	\langle \tilde{A}^{m_1}{}_{m_2}(q)\tilde{A}^{m_3}{}_{m_4}\rangle = \frac{N}{4\pi \lambda}\cot\left(\frac{\pi \lambda}{2}\right)\frac{(I_f)^{m_1m_3}_{m_2m_4}}{|q|}\,, \quad \langle \tilde{S}(q)\tilde{S}\rangle = \frac{NN_s}{4\pi\lambda}\cot\left(\frac{\pi \lambda}{2}\right)\frac{1}{|q|}\,.
\eeq
Following~\cite{GurAri:2012is} we find the three-point functions to be
\begin{align}
\begin{split}
\label{E:critferm3pt}
	\langle \tilde{A}^{m_1}{}_{m_2}(q)\tilde{A}^{m_3}{}_{m_4}(q')\tilde{A}^{m_5}{}_{m_6}\rangle & = -\frac{N}{2\pi\lambda} \left( \cot\left(\frac{\pi \lambda}{2}\right)+2\pi \lambda \tilde{\lambda}_{6F}'' \cot^3\left(\frac{\pi \lambda}{2}\right)\right) \frac{(H_f)^{m_1m_3m_5}_{m_2m_4m_6}}{|q||q'||q+q'|}\,,
	\\
	\langle \tilde{S}(q)\tilde{A}^{m_1}{}_{m_2}(q')\tilde{A}^{m_3}{}_{m_4}\rangle & = - \frac{N}{2\pi\lambda}\left( \cot\left(\frac{\pi\lambda}{2}\right) + 2\pi\lambda N_f\tilde{\lambda}_{6F}' \cot^3\left(\frac{\pi \lambda}{2}\right)\right) \frac{(I_f)^{m_1m_3}_{m_2m_4}}{|q||q'||q+q'|}\,,
	\\
	\langle \tilde{S}(q)\tilde{S}(q')\tilde{S}\rangle & = -\frac{NN_f}{2\pi\lambda}\left( \cot\left(\frac{\pi \lambda}{2}\right) + 2\pi \lambda N_f^2\tilde{\lambda}_{6F}\cot^3\left(\frac{\pi\lambda}{2}\right)\right)\frac{1}{|q||q'||q+q'|}\,.
\end{split}
\end{align}
	
%***********************************************
\subsubsection{Critical scalars and regular fermions}
%***********************************************

In this case all three-point functions of spin-zero bilinears are constants, and so contact terms. We do not consider them further.

%***********************************************
\section{Testing dualities}
%***********************************************

%***********************************************
\subsection{Critical bosons and regular fermions}
%***********************************************

Let us put the pieces together, beginning with the simplest theory, that of $U(N)_k$ Chern-Simons theory coupled to $N_s$ critical scalars and $N_f$ regular fermions. We found the free energy to be~\eqref{E:masterF2}
\begin{equation*}
	F = N_f F_f(N,\lambda,\mu_F) + N_s F_{s,c}(N,\lambda,\mu_{B,c})\,,
\end{equation*}
where $F_f$ is the free energy of the theory with a single regular fermion, and $F_{s,c}$ that of the theory of a single critical scalar. The fermion and boson thermal pole masses $\mu_F$ and $\mu_{B,c}$ are given by~\eqref{E:masterF1}.

This theory has scalar and fermion bilinears of spin-zero, $J_b$ and $J_f$, of dimension $2+O(1/N)$. Their three-point functions are pure contact terms. There are also spin $s=1,2,..$ operators $J_b^s$ and $J_f^s$ in the adjoint representation of $U(N_s)$ and $U(N_f)$ respectively, whose two- and three-point functions are essentially those of the spin $s=1,2,..$ bilinears of the theory with a single critical scalar, or a critical fermion. 

As a result the only non-trivial test of the master duality~\eqref{E:master} we can perform\footnote{It would be interesting to study insertions of the mixed bilinears $ (\bar{\psi}\phi)$ of dimension $\frac{3}{2}+O(1/N)$. The three-point function of two mixed bilinears with $J_b$ and $J_f$ would provide a non-trivial test.} is to see whether $F$ is invariant under 
\beq
	N \leftrightarrow |k|-N\,, \qquad \lambda \leftrightarrow \lambda-\text{sgn}(\lambda)\,, \qquad N_s \leftrightarrow N_f\,.
\eeq
This immediately follows from the known result~\cite{Aharony:2012ns} that the free energies $F_f$ and $F_s$ are exchanged under this map,
\beq
	F_f(N,\lambda,\mu) = F_{s,c}(|k|-N,\lambda - \text{sgn}(\lambda),\mu)\,, 
\eeq
and further that the equations for the pole masses,~\eqref{E:masterF1}, which do not depend on $N_f$ or $N_s$, are exchanged under it.

%***********************************************
\subsection{Regular scalars, critical fermions}
\label{S:testRC}
%***********************************************

The thermal free energy of $U(N)_k$ Chern-Simons theory coupled to $N_s$ regular scalars is given by~\eqref{E:Fregreg},
\begin{equation*}
	F_{reg} = -\frac{NN_sV_2}{2\pi^2i \beta^3|\lambda|}\left\{ \frac{\mu_B^2}{3}\text{Li}_2\left( e^{-\mu_B +\pi i |\lambda|}\right) + \int_{\mu_B}^{\infty} dy \,y\, \text{Li}_2\left( e^{-y-\pi i |\lambda|}\right) - \text{c.c.}\right\}\,,
\end{equation*}
where the thermal pole mass $\mu_B^2$ is determined by~\eqref{E:finalFormEOMs}
\begin{equation}
\label{E:eomregscalar}
	\mu_B= \frac{1}{2}\sqrt{1+\frac{N_s^2\tilde{\lambda}_6}{8\pi^2\lambda^2}}\left( |\lambda|\mu_B - \frac{1}{\pi i} \left[ \text{Li}_2\left(e^{-\mu_B+\pi i |\lambda|}\right)-\text{c.c.}\right]\right)\,,
\end{equation}
The two- and three-point functions of the scalar bilinears $S_b$ and $A_b$ are given in Subsections~\ref{S:twopoint} and~\ref{S:threepoint} respectively.

The thermal free energy of $U(N)_k$ Chern-Simons theory coupled to $N_f$ critical fermions is~\eqref{E:Fcritferm}, which we reprise
\begin{equation*}
	F_{crit} = \frac{NN_fV_2}{2\pi^2i \beta^3|\lambda|}\left\{ \frac{\mu_{F,c}^2}{3}\text{Li}_2\left( -e^{-\mu_{F,c} +\pi i |\lambda|}\right) + \int_{\mu_{F,c}}^{\infty} dy \,y\, \text{Li}_2\left( -e^{-y+\pi i |\lambda|}\right) - \text{c.c.}\right\}\,,
\end{equation*}
where $\mu_{F,c}$ is the thermal pole mass given by~\eqref{E:Gcritferm} and~\eqref{E:eomcritferm},
\begin{equation}
\label{E:eomcritfermion}
	(1-|\lambda|+G)\mu_{F,c} = -\frac{1}{\pi i}\left[ \text{Li}_2\left(-e^{-\mu_{F,c}+\pi i |\lambda|}\right)-\text{c.c.}\right]\,, \quad G = (1-2\pi \lambda N_s^2 \tilde{\lambda}_{6F})^{-1/2}\,,
\end{equation}
where $\tilde{\lambda}_{6F}$ is the triple trace coupling $\sim \tilde{S}^3$ defined in~\eqref{E:critfermTripleTrace}. The two- and three-point functions of critical fermion ``bilinears'' $\tilde{S}$ and $\tilde{A}$ are given in Subsections~\ref{S:2ptcritferm} and~\ref{S:3ptcritferm}.

Comparing the two-point functions of scalar bilinears in the theory of $N_f$ regular scalars in~\eqref{E:twopointstarting} with those of the bilinears in the theory of $N_f$ critical fermions in~\eqref{E:critferm2pt}, we see that they are consistent with duality under
\begin{equation*}
	N\leftrightarrow |k|-N\,, \qquad k\to -k\,, \qquad S_b \leftrightarrow \pm \tilde{S}\,, \qquad A_b\leftrightarrow \pm \tilde{A}\,.
\end{equation*}

The three-point functions in the theory of regular scalars are given in~\eqref{E:3ptscalar2} and those in the theory of critical fermions are in~\eqref{E:critferm3pt}. Comparing the two, we see that they are also consistent provided that $S_b \leftrightarrow -\tilde{S}$, $A_b\leftrightarrow -\tilde{A}$, and
\beq
	z_6 \leftrightarrow -2\pi \lambda N_f^2 \tilde{\lambda}_{6F} \,, \qquad z_6' \leftrightarrow -2\pi \lambda N_f \tilde{\lambda}_{6F}' \,, \qquad z_6'' \leftrightarrow -2\pi \lambda \tilde{\lambda}_{6F}''\,.
\eeq
Solving for the transformations of $\tilde{\lambda}_6$, $\tilde{\lambda}_6'$, and $\tilde{\lambda}_6''$, we find essentially three copies of the same map
\begin{align}
\begin{split}
\label{E:critfermduality}
	\tilde{\lambda}_6 & \to \frac{8\pi^2}{N_f^2}(1-|\lambda|^2) (3-8\pi \lambda N_f^2 \tilde{\lambda}_{6F})\,,
	\\
	\tilde{\lambda}_6' & \to \frac{8\pi^2}{N_f}(1-|\lambda|^2)(3-8\pi \lambda N_f \tilde{\lambda}_{6F}')\,,
	\\
	\tilde{\lambda}_6'' & \to 8\pi^2(1-|\lambda|^2)(3-8\pi \lambda \tilde{\lambda}_{6F}'')\,.
\end{split}
\end{align}

In their study of the free energy of $U(N)_k$ Chern-Simons theory coupled to a single regular scalar or a single critical fermion, the authors of~\cite{Aharony:2012ns} showed that, if $\mu$ solves the equation for $\mu_B$,~\eqref{E:eomregscalar}, then it also solves the equation for $\mu_{F,c}$~\eqref{E:eomcritfermion} provided that we transform $\tilde{\lambda}_6$ as above (for $N_f=1$). This result continues to hold for $N_f>1$, since introducing $N_f$ amounts to a rescaling of $\tilde{\lambda}_6$ in~\eqref{E:eomregscalar} and of~$\tilde{\lambda}_{6F}$ in~\eqref{E:eomcritfermion}. It then follows that the free energy $F_{reg}$ of the theory with regular scalars is equal to that of the theory with critical fermions under $N\to |k|-N$, $k\to -k$, and the map of triple-trace couplings in~\eqref{E:critfermduality}.

%***********************************************
\subsection{Regular bosons and regular fermions}
\label{S:testRR}
%***********************************************

The thermal free energy is given in~\eqref{E:Fregreg}, and it takes the form
\beq
\label{E:Fregreg2}
	F = N_f F_f(N,\lambda,\mu_F) + N_s F_s(N,\lambda,\mu_B)\,,
\eeq
where $F_f$ and $F_b$ are the thermal free energies of $U(N)_k$ Chern-Simons theory coupled to a single regular fermion or boson. They satisfy
\beq
\label{E:Fmapregreg}
	F_f(N,k,\mu) = F_s(|k|-N,\lambda-\text{sgn}(\lambda),\mu)\,.
\eeq
The thermal pole masses $\mu_B$ and $\mu_F$ obey~\eqref{E:finalFormEOMs}, which we reproduce here
\begin{align}
\begin{split}
\label{E:betterEOMs}
	\mu_B^2 & = (z_6+1) h_B^2 +\tilde{x}_4 (h_F - 2 \tilde{x}_4 h_B)h_F\,,
	\\
	\mu_F & = h_F - \tilde{x}_4 h_B\,,
\end{split}
\end{align}
while $h_B$ and $h_F$ are given by~\eqref{E:hBhF}, $z_6$ was defined in~\eqref{E:y}, and
\beq
	\tilde{x}_4 = N_f x_4 = \frac{N_f \tilde{\lambda}_4}{4\pi \lambda} = \frac{N_f \lambda_4 + \lambda_4'}{4\pi\lambda}\,,
\eeq
is the normalized double-trace coupling $\sim S_b S_f$. The theory has spin-zero operators $J_b$ and $J_f$, whose two- and three-point functions are given in Subsections~\ref{S:2ptregreg} and~\ref{S:3ptregreg}.

In terms of the decomposition of the multi-trace couplings at the beginning of Section~\ref{S:correlators}, the free energy only depends on the double-trace coupling $\tilde{\lambda}_4 S_b S_f$ and triple-trace coupling $\tilde{\lambda}_6 S_b^3$.

As we remarked before, these results are consistent with a duality only when $N_s = N_f$. From the point of view of the thermal free energy, the duality exchanges the thermal pole masses $\mu_B \leftrightarrow \mu_F$ and $N_f \leftrightarrow N_s$. But, then the operators of the original theory and its putative dual do not match: the former contains a unique spin-zero operator $J_b= (\phi^{\dagger}\phi)$ of dimension $1+O(1/N)$ in the adjoint representation of $U(N_s)$, while the latter has a unique scalar $\tilde{J}_b=(\Phi^{\dagger}\Phi)$ of dimension $1+O(1/N)$, but in the adjoint representation of $U(N_f)$. 

So we must have $N_s = N_f$. The marginal couplings of this model can be tuned to a unique point with enhanced $\mathcal{N}=2$ SUSY, and there is strong evidence~\cite{Giveon:2008zn,Benini:2011mf} that this theory is self-dual. The duality map takes the marginal deformations of this theory to those of the dual, and so at large $N$ we expect there to be a valid duality map at least in a neighborhood of the $\mathcal{N}=2$ point.

%***********************************************
\subsubsection{$\mathcal{N}=2$ Giveon-Kutasov duality}
\label{S:N=2}
%***********************************************

Consider the $\mathcal{N}=2$ point. The triple-trace and double-trace couplings are fixed as in~\eqref{E:N=2couplings}. These interactions then take the iterated form
\beq
\label{E:N=2potential}
	S_{int} = \int d^3x \left(\frac{4\pi^2\lambda^2}{N^2}( \phi^{\dagger}_{m}\phi^{n})(\phi^{\dagger}_{n}\phi^p)(\phi^{\dagger}_p \phi^m)+ \frac{4\pi \lambda}{N} (\phi^{\dagger}_m \phi^n)(\bar{\psi}_n\psi^m)  + \hdots\right)\,,
\eeq
where the dots indicate terms that are subleading at large $N$. Equivalently, the normalized multi-trace couplings $\tilde{x}_4$, $x_4'$, $z_6$, $z_6'$, and $z_6''$ that we introduced in the last Section are
\beq
	\tilde{x}_4 = x_4' = 1\,, \qquad z_6 = z_6' = z_6'' = 0\,.
\eeq

This model is expected to be self-dual under $N \to |k|-N$ and $\lambda\to\lambda-\text{sgn}(\lambda)$. At finite temperature, the equations for the pole masses~\eqref{E:betterEOMs} become
\beq
	\mu_B^2 = (h_B-h_F)^2 = \mu_F^2=\mu^2\,,
\eeq
and the equal bosonic and fermionic pole masses satisfy
\beq
	\mu = \frac{1}{\pi i}\left[ \text{Li}_2\left( e^{-\mu + \pi i |\lambda|}\right) +\text{Li}_2\left(-e^{-\mu-\pi i |\lambda|}\right)-\text{c.c.}\right]\,.
\eeq
The free energy is then manifestly self-dual on account of~\eqref{E:Fmapregreg}, and simplifies to
\beq
	F = N_f F_{\chi}(N,\lambda) = N_f(F_f(N,\lambda,\mu) + F_s(N,\lambda,\mu))\,.
\eeq
Here $F_{\chi}(N,\lambda)$ is the free energy of the $\mathcal{N}=2$ theory of $U(N)_k$ Chern-Simons theory coupled to a single chiral multiplet in the fundamental representation. Stated simply, we then have
\beq
\label{E:N=2selfduality}
	F_{\chi}(N,\lambda) = F_{\chi}(|k|-N,\lambda-\text{sgn}(\lambda))\,.
\eeq

The key feature here is that the matter interactions in~\eqref{E:N=2potential} iterate. That feature also holds for $U(N)_k$ Chern-Simons theory coupled to $N_f$ fundamental chiral multiplets $(\phi^m,\psi^m)$ and $\overline{N}_f$ antifundamental chiral multiplets $(\Phi^\alpha,\Psi^{\alpha})$, whose self-interactions are given by
\begin{align}
\begin{split}
	S_{int} = \int d^3x & \left( \frac{4\pi^2}{k^2} (\phi^{\dagger}_m\phi^n)(\phi^{\dagger}_n\phi^p)(\phi^{\dagger}_p \phi^m) + \frac{4\pi }{k}(\phi^{\dagger}_m\phi^n)(\bar{\psi}_n\psi^m) + \hdots\right.
		\\
	& \qquad \left.+ \frac{4\pi^2}{k^2}(\Phi^{\alpha}\Phi^{\dagger}_{\beta})(\Phi^{\beta}\Phi^{\dagger}_{\gamma})(\Phi^{\gamma}\Phi^{\dagger}_{\alpha}) + \frac{4\pi }{k} (\Phi^{\alpha}\Phi^{\dagger}_{\beta})(\bar{\Psi}_{\alpha}\Psi^{\beta})+ \hdots\right)\,.
\end{split}
\end{align}
The free energy of this model follows from our analysis in Section~\ref{S:freeenergy}. It is given by
\beq
\label{E:FN=2splitting}
	F = N_f F_{\chi}(N,\lambda) + \overline{N}_f F_{\bar{\chi}}(N,\lambda)\,,
\eeq
where we are somewhat pedantically defining $F_{\bar{\chi}}(N,\lambda)$ to be the free energy of $\mathcal{N}=2$ $U(N)_k$ Chern-Simons theory coupled to an antifundamental chiral multiplet. That free energy equals $F_{\chi}$, but we split the free energy this way because we expect the free energy to continue to factorize even in the presence of mass deformations or chemical potentials for the global symmetry.

The more general dualities of~\cite{Benini:2011mf} equate this theory with a magnetic dual, at large $N$ given by $U(|k|-N)_{-k}$ Chern-Simons theory coupled to $N_f$ antifundamental chirals, $\overline{N}_f$ fundamental chirals, and $N_f\overline{N}_f$ gauge-neutral ``mesons.'' The mesons do not contribute at large $N$, and so we are in a position to test these dualities as well. However this test is trivial, on account of the splitting~\eqref{E:FN=2splitting} and the self-duality at $N_f=1$~\eqref{E:N=2selfduality}.

The two- and three-point functions of scalar and fermion bilinears also simplify dramatically at the $\mathcal{N}=2$ point of $U(N)_k$ Chern-Simons theory coupled to $N_f$ fundamental multiplets. From Subsection~\ref{S:2ptregreg} the two-point functions become the duality-invariant expressions
\begin{align}
\begin{split}
	\langle (A_b)^{m_1}{}_{m_2}(q) (A_b)^{m_3}{}_{m_4}\rangle & = \frac{N}{2\pi\lambda}\sin\left(\frac{\pi \lambda}{2}\right)\cos\left( \frac{\pi\lambda}{2}\right)\frac{(I_b)^{m_1m_3}_{m_2m_4} }{|q|}\,, 
	\\
	\langle S_b(q)S_b\rangle &= \frac{NN_f}{2\pi\lambda}\sin\left(\frac{\pi \lambda}{2}\right)\cos\left(\frac{\pi \lambda}{2}\right)\frac{1}{|q|}\,,
	\\
	\langle (A_f)^{m_1}{}_{m_2}(q) (A_f)^{m_3}{}_{m_4}\rangle & = -\frac{N}{2\pi\lambda}\sin\left(\frac{\pi\lambda}{2}\right)\cos\left(\frac{\pi\lambda}{2}\right)|q| (I_f)^{m_1m_3}_{m_2m_4}\,, 
	\\
	\langle S_f(q)S_f\rangle &= - \frac{NN_f}{2\pi\lambda}\sin\left(\frac{\pi\lambda}{2}\right)\cos\left(\frac{\pi\lambda}{2}\right)|q|\,.
\end{split}
\end{align}
The three-point functions similarly simplify, e.g.
\beq
	\langle S_b(q)S_b(q')S_b\rangle = \frac{N N_f}{8\pi\lambda}\frac{\sin(2\pi \lambda)}{|q||q'||q+q'|}\,.
\eeq

%***********************************************
\subsubsection{$\mathcal{N}<2$}
%***********************************************

Now consider a general value of the multi-trace couplings. As we will see this theory is self-dual under
\beq
	N \leftrightarrow |k|-N\,, \qquad \lambda\leftrightarrow \lambda-\text{sgn}(\lambda)\,,
\eeq
along with a suitable action on the multi-trace couplings. Rather than simply presenting the action, let us try to explain where it comes from. 

In order for the free energy~\eqref{E:Fregreg2} to be invariant under the duality map,~\eqref{E:Fmapregreg} implies that the duality exchanges bosonic and fermionic pole masses, $\mu_B \leftrightarrow \mu_F$. Under the combination of $|\lambda|\to 1-|\lambda|$ and $\mu_b\leftrightarrow\mu_F$, it follows from~\eqref{E:hBhF} that
\beq
	h_B \to \mu_F - h_F\,, \qquad h_F \to \mu_B - h_B\,,
\eeq
It must then be the case that after these transformations the equations~\eqref{E:betterEOMs} are left invariant, so that they are still solved by $\mu_B$ and $\mu_F$. The second equation for $\mu_F$ is left invariant provided that
\beq
	\tilde{x}_4 \to\frac{1}{\tilde{x}_4}\,.
\eeq
The other equation for $\mu_B$ is then left invariant provided that $z_6$ is mapped as
\beq
	z_6 \to- \frac{z_6}{\tilde{x}_4^3}\,.
\eeq
As a check, this map squares to the identity. So invariance of the free energy fixes the action of the duality map on $\tilde{\lambda}_4$ and $\tilde{\lambda}_6$. 

For $N_f = 1$, so that $\tilde{x}_4 = \frac{\lambda_4}{4\pi\lambda}$ and $z_6 =\frac{\hat{\lambda}^2}{4\lambda^2}-1$, this is the same duality map obtained in~\cite{Aharony:2012ns}, under which $\frac{\hat{\lambda}^2}{4\lambda^2}\to \frac{1}{\tilde{x}_4^3}\left(1+\tilde{x}_4^3-\frac{\hat{\lambda}^2}{4\lambda^2}\right)$. 

Is this duality map consistent with the two- and three-point functions of bilinears at zero temperature? And how do the other multi-trace couplings map under the duality? Let us begin to answer these questions by studying the two- and three-point functions of $S_b$, ~\eqref{E:twoptregreg1} and~\eqref{E:regreg3ptSSS}. Writing them out in full, they are
\begin{align}
\begin{split}
	\langle S_b(q)S_b\rangle &= \frac{NN_f}{4\pi\lambda} \frac{\tan\left(\frac{\pi \lambda}{2}\right)}{1+\tilde{x}_4^2\tan^2\left(\frac{\pi \lambda}{2}\right)} \frac{1}{|q|} = \frac{NN_f}{4\pi\lambda}\frac{1}{\tilde{x}_4(X+X^{-1})}\frac{1}{|q|}\,,
	\\
	\langle S_b(q)S_b(q')S_b\rangle& = \frac{NN_f}{2\pi\lambda}\frac{1}{\left(1+\tilde{x}_4^2 \tan^2\left(\frac{\pi \lambda}{2}\right)\right)^3}\frac{1}{|q||q'||q+q'|}
	\\
	& \qquad \qquad \times \left( \tan\left(\frac{\pi\lambda}{2}\right)-z_6 \tan^3\left(\frac{\pi \lambda}{2}\right)- \tilde{x}_4^3\tan^5\left(\frac{\pi \lambda}{2}\right)\right)
	\\
	& = \frac{N N_f}{2\pi\lambda} \frac{1}{(X+X^{-1})^3} \left( \frac{X^{-2}}{\tilde{x}_4} - \frac{z_6}{\tilde{x}_4^3 }- \frac{X^2}{\tilde{x}_4^2}\right)\,,
\end{split}
\end{align}
where the in the last line we have written the three-point function in a more suggestive form using $X = \tilde{x}_4 \tan\left(\frac{\pi\lambda}{2}\right)$. (Under the duality $X\to -X^{-1}$, $\tilde{x}_4\to \frac{1}{\tilde{x}_4}$, and $z_6 \to -\frac{z_6}{\tilde{x}_4^3}$.)
Under the duality map we have
\beq
	\langle S_b(q)S_b\rangle \to  \tilde{x}_4^2 \langle S_b(q)S_b\rangle\,,
\eeq
along with
\beq
	\langle S_b(q)S_b(q')S_b\rangle \to -\tilde{x}_4^3 \langle S_b(q)S_b(q')S_b\rangle\,.
\eeq
(The first and third terms in the three-point function are exchanged under the duality, with the middle term left invariant, all up to multiplication by $-\tilde{x}_4^3$.)
This is clearly consistent with the duality, and gives us its action on the operator $S_b = (\phi^{\dagger}_m\phi^m)$,
\beq
	S_b \to -\tilde{x}_4 S_b\,.
\eeq

Now consider the two-point function of the adjoint bilinear $A_b$,
\beq
	\langle (A_b)^{m_1}{}_{m_2}(q)(A_b)^{m_3}{}_{m_4}\rangle  = \frac{N}{4\pi\lambda}\frac{1}{x_4'(X'+X'^{-1})}\frac{(I_b)^{m_1m_3}_{m_2m_4}}{|q|} \,,
\eeq
with $X' = x_4' \tan\left(\frac{\pi\lambda}{2}\right)$. The three-point functions of $A_b$ with itself or $S_b$ similarly become
\begin{align}
\begin{split}
	\langle (A_b)^{m_1}{}_{m_2}(q) (A_b)^{m_3}{}_{m_4}(q')(A_b)^{m_5}{}_{m_6}\rangle & = \frac{N}{2\pi\lambda}\frac{1}{(X'+X'^{-1})^3}
	\\
	&\qquad \times \left(\frac{X'^{-2}}{x_4'}-\frac{z_6''}{x_4'^3}-\frac{X'^2}{x_4'^2}\right)\frac{(H_b)^{m_1m_3m_5}_{m_2m_4m_6}}{|q||q'||q+q'|}\,,
	\\
	\langle S_b(q)(A_b)^{m_1}{}_{m_2}(q)(A_b)^{m_3}{}_{m_4}\rangle & = \frac{N}{2\pi\lambda}\frac{1}{(X+X^{-1})(X'+X'^{-1})^2}
	\\
	& \qquad\times \left( \frac{X'^{-2}}{\tilde{x}_4} - \frac{z_6'}{\tilde{x}_4x_4'^2} - \frac{X'^2}{x_4'^2}\right)\frac{(I_b)^{m_1m_3}_{m_2m_4}}{|q||q'||q+q'|}\,.
\end{split}
\end{align}
These are consistent with the duality map provided that
\beq
	A_b \to -x_4' A_b\,, \qquad x_4' \to \frac{1}{x_4'} \,, \qquad z_6' \to -\frac{z_6'}{\tilde{x}_4 x_4'^2}\,, \qquad z_6'' \to -\frac{z_6''}{x_4'^3}\,.
\eeq
The transformation of $x_4'$ implies that $X'\to -1/X'$.

What of the fermion bilinears $S_f$ and $A_f$? Their two-point functions are
\begin{align}
\begin{split}
	\langle (A_f)^{m_1}{}_{m_2}(q)(A_f)^{m_3}{}_{m_4}\rangle & = -\frac{N}{4\pi\lambda}\frac{1}{x_4'(X'+X'^{-1})}|q| (I_f)^{m_1m_3}_{m_2m_4}\,,
	\\
	\langle S_f(q) S_f\rangle & = -\frac{NN_f}{4\pi \lambda}\frac{1}{\tilde{x}_4(X+X^{-1})}|q|\,,
\end{split}
\end{align}
and, as an example, the three-point function of $S_f$ with two $S_b$'s is
\beq
\label{E:SbSbSf}
	\langle S_b(q)S_b(q')S_f\rangle = \frac{NN_f}{2\pi\lambda}\frac{1}{(X+X^{-1})^3}\left( \frac{X^{-1}}{\tilde{x}_4} - \frac{Xz_6}{\tilde{x}_4^3} + \frac{X}{\tilde{x}_4^2}\right)\,.
\eeq
Under the duality map we have
\beq
\label{E:SfSftransform}
	\langle S_f(q) S_f\rangle \to \tilde{x}_4^2 \langle S_f(q) S_f\rangle\,,
\eeq
which would suggest that $S_f\to \pm \tilde{x}_4S_f$, but $\langle S_b(q)S_b(q')S_f\rangle$ does not map to a constant times the original three-point function. The culprit is the second term in~\eqref{E:SbSbSf}. As a result the three-point function is inconsistent with duality, at least under the map $S_f \to \pm \tilde{x}_4S_f$. 

The resolution to this seeming breakdown of duality\footnote{We are grateful to O.~Aharony for pointing out this resolution.} is that the duality transformation of $S_f$ may involve operator mixing, since $S_f$, $S_b^2$, and $\text{tr}(A_b^2)$ are all of dimension $2+O(1/N)$. 

We fix the $c_i$ as follows. Let
\beq
	S'_f =  c_0 S_f +\frac{c_1}{NN_f}S_b^2 + \frac{c_2}{N}\text{tr}(A_b^2)\,,
\eeq
be the transformed fermion bilinear. Large $N$ factorization implies that to leading order in large $N$,
\beq
	\langle S_b(q)S_b(q') S_b^2\rangle  = \frac{N^2N_f^2}{8\pi^2\lambda^2}\frac{1}{\tilde{x}_4^2(X+X^{-1})^2}\frac{1}{|q||q'|}\,, \qquad \langle S_b(q)S_b(q')\text{tr}(A_b^2)\rangle = 0\,.
\eeq
Thus
\beq
	\langle S_f'(q)S'_f\rangle = c_0^2 \langle S_f(q)S_f\rangle + \hdots\,,
\eeq
where the dots indicate subleading terms at large $N$. Comparing with~\eqref{E:SfSftransform} we see that $c_0 = \pm \tilde{x}_4$. The three-point function with $S'_b = -\tilde{x}_4 S_b$ is 
\begin{align}
	\langle S_b'(q)S_b'(q')S_b'\rangle &= \tilde{x}_4^2c_0 \langle S_b(q)S_b(q')S_f\rangle + \frac{\tilde{x}_4^2 c_1}{N}\langle S_b(q)S_b(q')S_b^2\rangle
	\\
	\nonumber
	& = \pm \tilde{x}_4^3\frac{NN_f}{2\pi\lambda}  \frac{1}{(X+X^{-1})^3} \left( \frac{X^{-1}}{\tilde{x}_4} - \frac{X z_6}{\tilde{x}_4^3} + \frac{X}{\tilde{x}_4^2} \pm \frac{c_1}{4\pi\lambda\,\tilde{x}_4^3}(X+X^{-1})\right)\frac{1}{|q||q'|}\,.
\end{align}
Under the duality that three-point function~\eqref{E:SbSbSf} transforms as
\beq
	\langle S_b(q)S_b(q')S_f\rangle \to -\tilde{x}_4^3\frac{NN_f}{2\pi\lambda}\frac{1}{(X+X^{-1})^3} \left( \frac{X^{-1}}{\tilde{x}_4} + \frac{X^{-1}z_6}{\tilde{x}_4^3} + \frac{X}{\tilde{x}_4^2}\right)\frac{1}{|q||q'|}\,.
\eeq
Comparing these two expressions we see that
\beq
	c_0 = -\tilde{x}_4\,, \qquad c_1 = -4\pi \lambda\, z_6\,.
\eeq
A similar computation involving the three-point function 
\beq
	\langle (A_b)^{m_1}{}_{m_2}(q)(A_b)^{m_3}{}_{m_4}(q')S_f\rangle = \frac{N}{2\pi\lambda}\frac{1}{(X'+X'^{-1})^2(X+X^{-1})} \left( \frac{X'^{-1}}{x_4'} - \frac{Xz_6'}{\tilde{x}_4x_4'^2} + \frac{X'}{x_4'^2}\right)\frac{1}{|q||q'|}\,,
\eeq
along with $\langle A_b(q)A_b(q')A_b^2\rangle$ fixes
\beq
	c_2 = -4\pi \lambda z_6'\,.
\eeq

By a similar analysis we find that the two-point function of $A_f$ and the mixed three-point functions with a single $A_f$ are consistent with duality under the map
\beq
	A_f \to -x_4' A_f - \frac{4\pi \lambda z_6''}{N}A_b^2 - \frac{4\pi \lambda z_6'}{NN_f}A_b S_b\,.
\eeq
All in all, we have four three-point functions involving the fermion bilinears $S_f$ and $A_f$, we use those to fix the four mixing parameters allowed by symmetry in their duality transformation.

%***********************************************
\subsubsection{Summary}
%***********************************************

Let us summarize the duality map. We can parameterize the matter interactions of $U(N)_k$ Chern-Simons theory coupled to $N_f$ regular scalars and fermions as
\begin{align}
\begin{split}
	S_{matter} = \int d^3x &\left\{ \bar{\psi}_m \slashed{D}\psi^m+|D\phi^m|^2 + \frac{4\pi \tilde{x}_4}{kN_f} S_b S_f  + \frac{4\pi x_4'}{k}\text{tr}(A_bA_f)\right.
	\\
	&\left.+\frac{4\pi^2(3+4z_6)}{3k^2N_f^2}S_b^3 + \frac{4\pi^2(3+4z_6')}{k^2N_f}S_b \text{tr}(A_b^2)+\frac{4\pi^2(3+4z_6'')}{3k^2}\text{tr}(A_b^3)\right\}\,,
\end{split}
\end{align}
up to terms which do not contribute at large $N$. We have obtained evidence that this model is dual under
\begin{equation*}
	N \leftrightarrow |k|-N\,, \qquad k\to -k\,,
\end{equation*}
along with
\beq
	\tilde{x}_4 \to \frac{1}{\tilde{x}_4}\,, \quad x_4'\to \frac{1}{x_4'} \,, \quad z_6 \to -\frac{z_6}{\tilde{x}_4^3}\,, \quad z_6' \to -\frac{z_6'}{\tilde{x}_4x_4'^2}\,, \quad z_6'' \to -\frac{z_6''}{x_4'^3}\,.
\eeq
The spin-zero scalar bilinears map as
\beq
	S_b \to -\tilde{x}_4 S_b\,, \quad A_b \to -\tilde{x}_4A_b\,,
\eeq
and the spin-zero fermion bilinears as
\beq
	S_f \to -\tilde{x}_4 S_f - \frac{4\pi}{k}\left(\frac{z_6}{N_f} S_b^2 + z_6' \text{tr}(A_b^2)\right)\,, \quad A_f \to -x_4'A_f -\frac{4\pi}{k}\left(  \frac{z_6'}{N_f}A_b S_f+z_6'' A_b^2\right)\,.
\eeq
A nice consistency check on our result is that the action of the duality squares to the identity on these couplings and operators.

We can compare our results with those recently obtained for $\mathcal{N}=1$ Chern-Simons-matter theory~\cite{Aharony:2019mbc}. This theory has an $\mathcal{N}=1$ submanifold parameterized by
\begin{align}
\begin{split}
	\tilde{x}_4 &= \frac{\omega_1+1}{2}\,, \quad x_4' = \frac{\omega_2+1}{2}\,, \quad z_6 = \frac{3}{4}\left( \omega_1^2-1\right)\,, 
	\\
	z_6'.&= \frac{1}{4} \left( \omega_2(2\omega_1+\omega_2)-3\right)\,, \quad\,\,\,\, z_6'' = \frac{3}{4}\left(\omega_2^2-1\right)\,,
\end{split}
\end{align}
where the parameters $\omega_1$ and $\omega_2$ are related to the quartic superpotential couplings $w_1$ and $w_2$ in~\eqref{E:supoCouplings} as
\beq
	w_1 = \frac{2\pi\lambda}{N_f}(\omega_1-\omega_2)\,, \qquad w_2 = 2\pi\lambda\,\omega_2\,.
\eeq
The $\mathcal{N}=2$ point is $\omega_1= \omega_2=1$. Curiously, $J_f$ does not mix with $J_b^2$ at that point. The duality above implies that the superpotential couplings $\omega_1$ and $\omega_2$ map as
\beq
	\omega_1 \to \frac{3-\omega_1}{1+\omega_1} \,, \qquad	\omega_2 \to \frac{3-\omega_2}{1+\omega_2}\,.
\eeq
This agrees with the result obtained in~\cite{Aharony:2019mbc} by looking at a subset of two- and three-point functions in the conformal theory, as well as the pole masses of the mass-deformed model. 

\acknowledgments

We would like to thank A.~Karch, S.~Minwalla, and especially O.~Aharony for useful discussions. KJ would also like to thank the Kavli Institute for Theoretical Physics for their support, where a portion of this work was completed. KJ was supported in part by the US Department of Energy under grant number DE-SC 0013682, and by the National Science Foundation under Grant No. NSF PHY-1748958.

\bibliographystyle{JHEP}
\bibliography{refs}

\end{document}